\documentclass[fleqn,usenatbib]{mnras}

\usepackage{newtxtext,newtxmath}

\usepackage[T1]{fontenc}
\usepackage{ae,aecompl}

\usepackage{graphicx}	
\usepackage{amsmath}	
\usepackage{amssymb}	
\usepackage{lipsum}

\newcommand{\etacar}{$\eta$ Car} 
\newcommand{\mpers}{\,\rm{ms}^{-1}} 
\newcommand{\kmpers}{\,\rm{\,km \, s^{-1}}} 
\newcommand{\au}{\,\rm{AU}} 
\newcommand{\kpc}{\,\rm{kpc}} 
\newcommand{\days}{\,\rm{days}} 
\newcommand{\solarl}{\,\rm{L_{\odot}}} 
\newcommand{\solarm}{\,\rm{M_{\odot}}} 
\newcommand{\solarr}{\,\rm{R_{\odot}}} 
\newcommand{\masslossrate}{\,\rm{M_{\odot} yr^{-1}}} 

\title[Orbits of powerful wind binaries]{Uncovering the orbital dynamics of stars hidden inside their powerful winds: application to $\eta$ Carinae and\newline RMC 140}

\author[D. Grant et al.]{
David Grant,$^{1}$\thanks{E-mail: david.grant@physics.ox.ac.uk}
Katherine Blundell$^{1}$
and James Matthews$^{2,1}$
\\
$^{1}$University of Oxford, Department of Physics, Keble Road, Oxford, OX1 3RH, U.K\\
$^{2}$Institute of Astronomy, University of Cambridge, Madingley Road, Cambridge, CB3 0HA, U.K\\
}

\date{Accepted 2020 March 5. Received 2020 February 11; in original form 2019 November 21}

\pubyear{2019}

\begin{document}
\label{firstpage}
\pagerange{\pageref{firstpage}--\pageref{lastpage}}
\maketitle

\begin{abstract}
Determining accurate orbits of binary stars with powerful winds is challenging. The dense outflows increase the effective photospheric radius, precluding direct observation of the Keplerian motion; instead the observables are broad lines emitted over large radii in the stellar wind. Our analysis reveals strong, systematic discrepancies between the radial velocities extracted from different spectral lines: the more extended a line's emission region, the greater the departure from the true orbital motion. To overcome these challenges, we formulate a novel semi-analytical model which encapsulates both the star's orbital motion and the propagation of the wind. The model encodes the integrated velocity field of the out-flowing gas in terms of a convolution of past motion due to the finite flow speed of the wind. We test this model on two binary systems. (1), for the extreme case $\eta$ Carinae, in which the effects are most prominent, we are able to fit the model to 10 Balmer lines from H-alpha to H-kappa concurrently with a single set of orbital parameters: time of periastron $T_{0}=2454848$ (JD), eccentricity $e=0.91$, semi-amplitude $k=69 \kmpers$ and longitude of periastron $\omega=241^\circ$. (2) for a more typical case, the Wolf-Rayet star in RMC 140, we demonstrate that for commonly used lines, such as \ion{He}{II} and \ion{N}{III/IV/V}, we expect deviations between the Keplerian orbit and the predicted radial velocities. Our study indicates that corrective modelling, such as presented here, is necessary in order to identify a consistent set of orbital parameters, independent of the emission line used, especially for future high accuracy work.
\end{abstract}




\section{Introduction}
\label{sec:introduction}

The role of dynamics and mass loss is interconnected with the evolutionary tracks of high-mass stellar binaries: during their lifetimes as stars to their eventual fate as supernovae and potential gravitational wave mergers \citep[][]{Sana2012BinaryStars, Smith2014MassStars, Belczynski2002AProperties}. Binary systems that include stars exhibiting powerful winds are therefore pertinent case studies for our overall picture of massive stars.

Powerful stellar winds are a feature of both cool (RSG, Extreme RSG, YSG, YHG)\footnotemark \footnotetext{Red supergiant (RSG), yellow supergiant (YSG), yellow hypergiant (YHG).\label{cool_stellar_types}} and hot (BSG, LBV, WR, Be, B[e])\footnotemark{} stellar types. Observationally these systems show spectral lines predominantly in emission owing to their extended atmospheres and dense winds where the lines are formed \citep[][]{Beals1929OnEmission}. This in turn presents certain diagnostic challenges, especially in the most extreme mass-loss cases where the outflows can obscure measurements of the central star. The optically-thick gas enshrouding the star moves the effective photospheric radius outwards in the flow; and therefore, deducing the obscured system's parameters can be problematic.
\footnotetext{Blue supergiant (BSG), luminous blue variable (LBV), Wolf-Rayet star (WR).\label{hot_stellar_types}}

The orbital parameters of stars are typically determined by fitting Keplerian models to radial velocity measurements, and the radial velocities are calculated from the Doppler shifts extracted from spectral lines. For stars without strong outflows, having absorption line spectra, the orbital motion can be tracked at the stellar surface. Precision on the order of ${\sim} 10 \mpers$ is achievable with state-of-the-art instrumentation and extraction algorithms \citep[][]{Konacki2005PrecisionCell, Konacki2010High-precisionHD210027}, nearing the magnitude at which intrinsic stellar variability becomes important \citep[][]{Saar1998MagneticSurvey}. 

For stars with powerful winds the emission-line spectra are manifestly more complex and systematic effects are the dominant source of error. The broad emission lines are formed far out in the stellar wind \citep[][]{Beals1929OnEmission}, encoding dynamical information from an extensive volume of outflowing gas. The winds show stratification in the line emission regions dependent on each line's excitation energy \citep[][]{Bowen1928THENEBULAE, Kuhi1973Wolf-RayetStratification, Willis1982P-CygniStars} and opacity \citep[][]{Hillier1987An50896}.

To overcome these challenges earlier work has focused on extracting radial velocities from high excitation lines, known to form deep down in the wind, such as \ion{He}{I/II} or \ion{N}{III/IV/V}, depending on the spectral type and availability of lines \citep[eg.][]{Shenar2019TheEvolution}. But, even for high excitation lines it is unclear whether the entire line emission region is co-moving along the star's orbital path. In fact, large discrepancies have been found between the radial velocities extracted from various high-energy lines in WR binaries \citep[][]{Stickland1984UltravioletCephei, Shylaja1986SpectrophotometricCephei}. 

It has been suggested that these discrepancies are the result of non-Keplerian velocities present in the extended wind \citep[][]{Shylaja1987TheBinaries}. Radial velocities may also be affected by the wind and radiation of a companion star: the formation of colliding winds can complicate the line profiles \citep[eg.][]{Marchenko2003The140, Groh2012OnSpectra, Groh2012ACarinae} and the radiation field can influence the ionisation structure of the primary star's wind \citep[][]{Kruip2011ConnectingAlgorithm, Madura2012ConstrainingEmission, Clementel20153DApastron, Clementel2015}. Consequently, emission lines may contain velocity information which is not purely Keplerian, even if the motion of the stars from which they arise is Keplerian. The degree to which extracted radial velocities deviate from the true Keplerian orbits will depend on the exact nature of the line transition and outflows in which they are emitted.

\subsection{\texorpdfstring{$\eta$}{} Carinae: the extreme case}
\label{subsec:intro_eta_cariane}
When it comes to testing models of lines formed in extended regions the LBV Eta Carinae (hereafter \etacar{}) is an extreme case study. Characteristic of LBVs, \etacar{} has an enormous mass-loss rate and relatively slow wind resulting in the longest wind-flow times to the emission-line regions of almost any emission-line star known today. As a consequence, the observational effects will be at their most pronounced in this system.

\etacar{} is located in the Trumpler 16 cluster in the Carina nebula \citep[][]{Walborn1973SomeComplex}. Situated at a distance of $2.3 \pm 0.1 \kpc$ \citep[][]{Allen1993TheCarinae, Smith2006TheCarinae} and generating a luminosity of  $L=5 \times 10^6 \solarl$ \citep[][]{Davidson1997ETAENVIRONMENT} it is one of the most luminous sources in our Galaxy. \etacar{} first captured the attention of astronomers due to its ``Giant Eruption" in the 1840s, where it is estimated to have ejected in excess of $12 \solarm$ of material over about a decade \citep[][]{Smith2003MassCarinae}, and formed the bipolar Homunculus nebula seen today \citep[][]{Currie1996AstrometricTelescope}. Further quasi-periodic eruptions have been detected, such as the so-called Lesser Eruption in 1890  \citep[][]{Walborn1977TheNebula, Humphreys1999EtaVariables} and a photometric event in 1941 \citep[][]{deVaucouleurs1952TheCarinae}. These outbursts are responsible for a complex environment of debris surrounding the source, including peculiar objects such as the ``Weigelt blobs" or ``speckle objects" \citep[][]{Weigelt1986EtaInterferometry, Dorland2004Did1941}.

\etacar{} is a binary system with a stable orbital period of $P=2022.7 \pm 1.3 \days{}$ \citep[][]{Damineli2008TheEvents}. The binary scenario was originally proposed by \citet[][1992 periastron data]{Damineli1997EtaBinary} and since that time the orbital parameters have been computed using numerous techniques and observational data. Overall, the general consensus is for a highly eccentric (${\sim} 0.9$) binary which is supported by models of X-ray light curves \citep[][]{Pittard1998TheCarinae, Corcoran2001TheLoss, Okazaki2008ModellingCollision} and \ion{He}{I/II} line equivalent widths \citep[][]{Richardson2015TheEvent, Teodoro2016HeMINIMA}. However, the exact orbital solution has been hard to establish by virtue of the differences in results obtained from various species and the influence of the companion \citep[][]{Groh2012OnSpectra}.

The primary star is an LBV with an extremely powerful wind: a mass-loss rate of $8.5 \times 10^{-4} \masslossrate$ and terminal wind velocity of $420 \kmpers$ \citep[][]{Hillier2001, Groh2012OnSpectra, Clementel20153DApastron}. Under these conditions, \citet[][]{Hillier2001} showed that H-alpha could be formed in an extended region between $6-60 \au$ (assuming $R_{\ast} = 60 \solarr$) outwards from the primary, and these results have been observationally supported by \citet[][]{Wu2017ResolvingCarinae}. As a result, we know the primary star displays spectral lines formed in highly extended regions from the central star.

Details of the companion are far more uncertain because it has never been directly observed; it is a few orders of magnitude less bright than the primary in visible and longer wavelengths, and is not seen in the UV either \citep[][]{Hillier2006TheCarinae}. Estimates for the companion's wind have been made from modelling X-ray observations, yielding estimates of a mass-loss rate of ${\sim} 10^{-5} \masslossrate$ and terminal wind velocity of ${\sim} 2000-3000 \kmpers$ \citep[][]{Pittard2002, Parkin2011SpiralingCarinae}.

\subsection{RMC 140: the typical case}
\label{subsec:intro_r140}
We also investigate RMC 140 (alternatively BAT99 103 or VFTS 509): a more typical emission-line star system. The aim here is to gauge how large the observable effects may be for lines formed in extended regions, for a system more representative of a larger population of stars, and the ramifications for their calculated orbital parameters.

RMC 140 is located in the Large Magellanic Cloud (LMC) and at a distance of $49.97 \pm 01.11 \kpc$ \citep[][]{Pietrzynski2013AnCent}. \citet[][]{Moffat1987TheSurroundings} first discovered RMC 140 is a binary system, finding a period of $2.76 \days{}$. The system is comprised of an O-star (type O4 V, $L=5.62 \times 10^5 \solarl$, hereafter RMC 140a) and a WR star (type WN5, $L=1.23 \times 10^6 \solarl$, hereafter RMC 140b) \citep[][]{Evans2011TheOverview, Shenar2019TheEvolution}. For our analysis we only concern ourselves with the line emission from RMC 140b due to its powerful WR wind; it has a mass-loss rate of $1.26 \times 10^{-5} \masslossrate$ and terminal wind velocity of $1300 \kmpers$ \citep[][]{Shenar2019TheEvolution}.
\newline

\begin{figure*}
	\includegraphics[width=\textwidth]{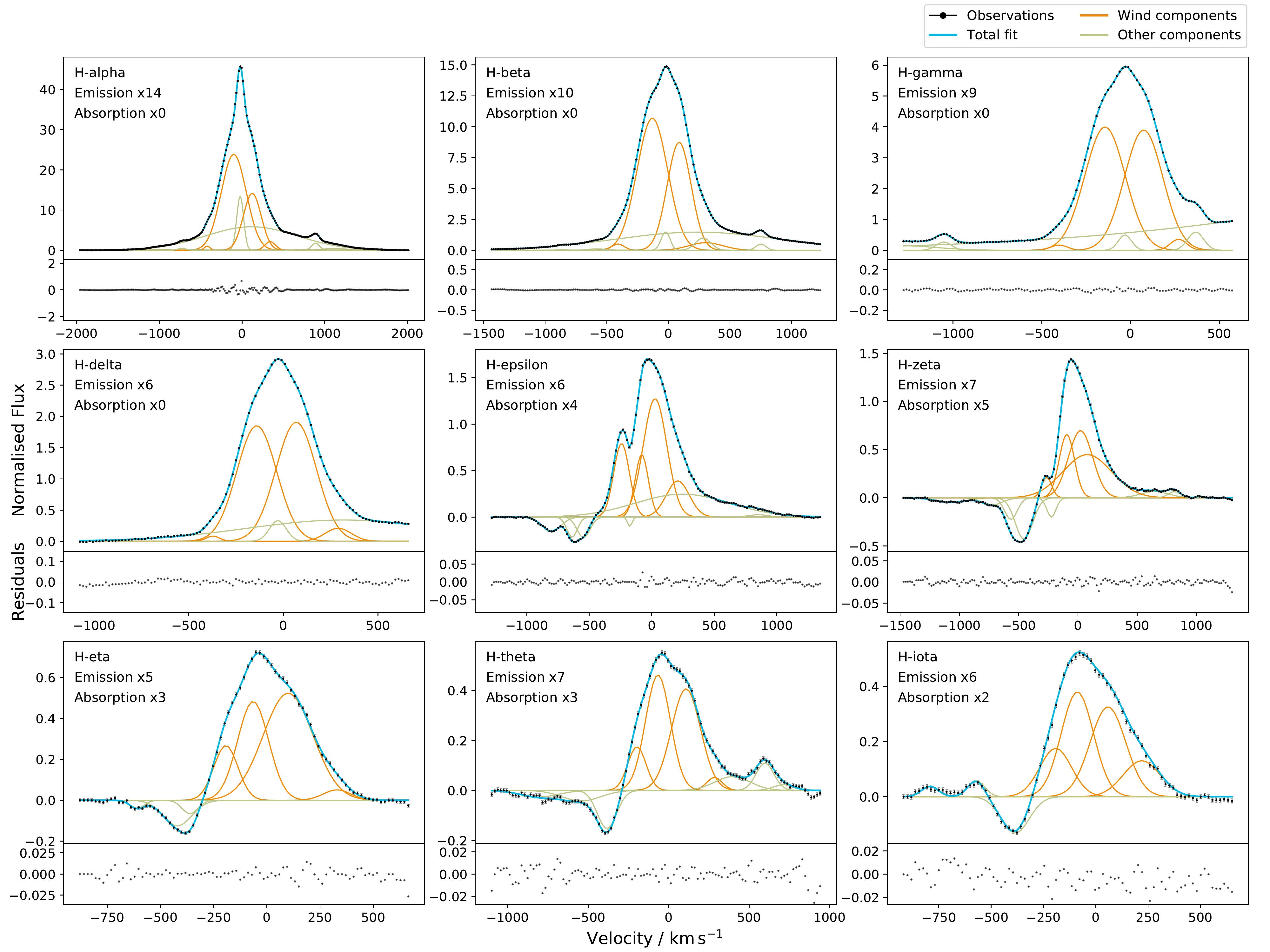}
    \caption{Multi-Gaussian fitting of simultaneously observed Balmer series spectral profiles near to apastron. Each panel indicates the Balmer line transition, the number of emission components (positive Gaussians) and the number of absorption components (negative Gaussians). The top of each panel shows the total model (light blue) and its Gaussian components (light green and orange) vs the observations (black). The Gaussian components attributed to wind emission are highlighted orange. The bottom of each panel shows the residuals of each fit.}
    \label{fig:multig_balmer_fits_apastron}
\end{figure*}

\begin{figure*}
	\includegraphics[width=\textwidth]{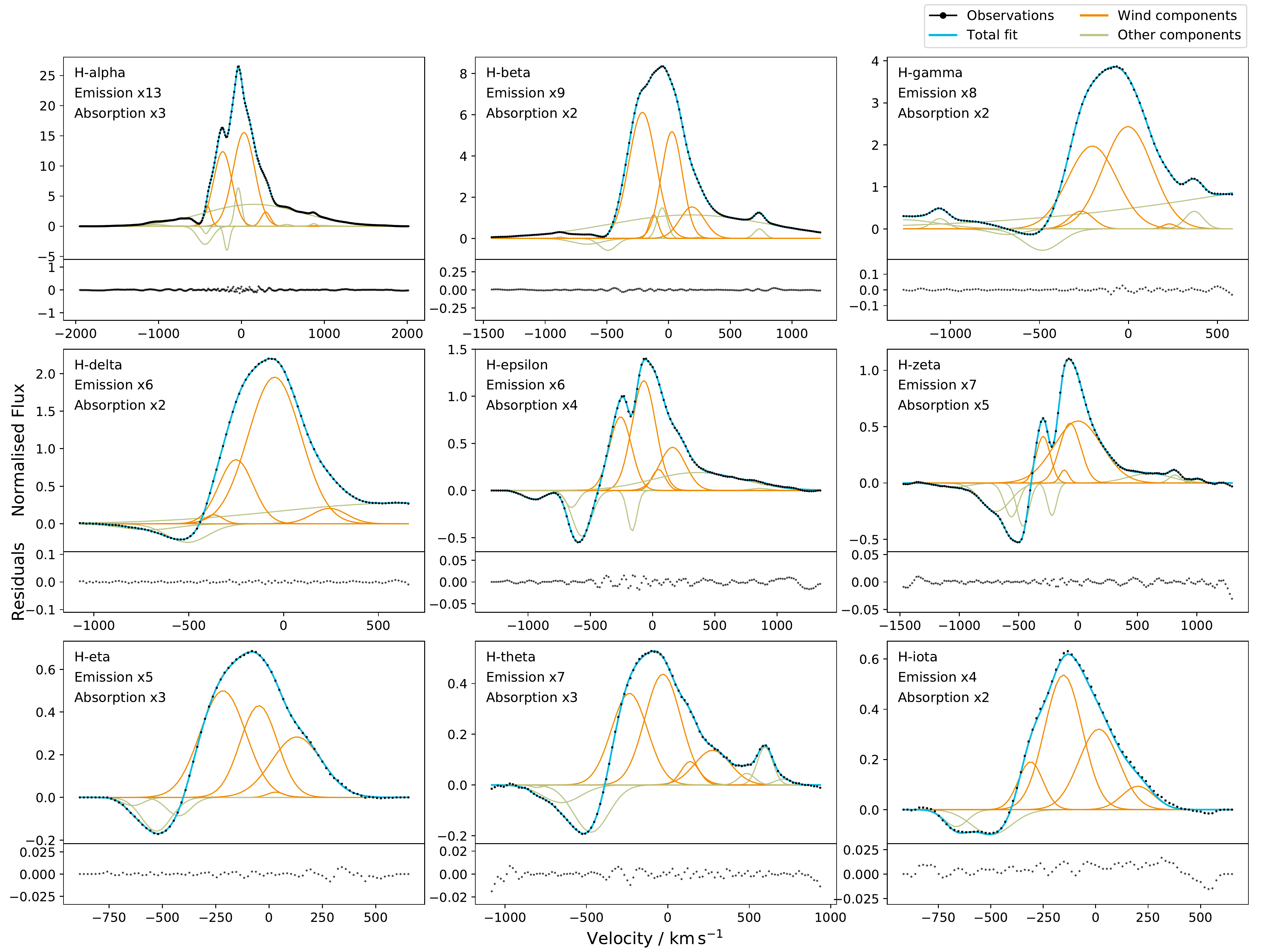}
    \caption{The same as Figure \ref{fig:multig_balmer_fits_apastron}, but for periastron.}
    \label{fig:multig_balmer_fits_periastron}
\end{figure*}

In this study we investigate the dynamical information encoded in stellar emission lines, principally for stars with powerful winds where line emission can be highly extended. \etacar{}, as previously introduced in Section \ref{subsec:intro_eta_cariane}, is our main case study. We present the observations and extraction methods used for computing radial velocities for \etacar{} in Section \ref{sec:observations}. In Section \ref{sec:modelling} we begin by fitting the radial velocity curves with Keplerian models and show how selecting different Balmer emission lines can result in changes to the resulting orbital parameters. We then describe a new semi-analytical model which aims to reconcile different emission lines' radial velocity curves with a single set of orbital parameters, and apply it to \etacar{}. In Section \ref{sec:discussion} we discuss several complexities of the model and the newly-derived orbital parameters for \etacar{}. We then apply our model to RMC 140, as previously introduced in Section \ref{subsec:intro_eta_cariane}, to understand how our model affects the interpretation of other emission-line stars' radial velocity curves. Finally, in Section \ref{sec:summary_and_conclusions} we summarise our findings.

\section{Observations}
\label{sec:observations}

The observations of \etacar{} used throughout this study are from an online open-source data archive\footnote{\url{http://etacar.umn.edu/archive/}}. We make use of data observed by the Gemini-South Multi-Object Spectrograph (GMOS) on the Gemini-South telescope. The dataset contains 310 individual spectra over 33 nights of observing. The dataset spans the 2009 periastron of \etacar{}, starting in June 2007 and ending in January 2010. The spectra are particularly well sampled at times near periastron, having 252 spectra in the phase interval $0.95<\phi<1.05$. The spectra cover a wavelength range of $3557<\lambda<7546$. The exposure times range from 1--180 seconds to optimise the signal-to-noise for different emission lines and instrumental configurations, and the resolving power varies in the interval $3700<R<4400$. For further technical information see the original work by \citet[][]{Mehner2011CriticalEvent} detailing the observing campaign.

The \etacar{} spectra contain a wealth of information entangled in a multitude of complex line profiles. Given the complexity of the system, the spectra contain light from the central star, companion, extended wind, Weigelt blobs, and scattering from dust in the Homunculus nebula. We are particularly interested in the lines that form in the extended wind of the primary. These lines are known to show variability reflecting the orbital cycles as extensively discussed by previous authors: the \ion{He}{I} lines \citep[][]{Damineli1997EtaBinary}, the \ion{He}{II} lines \citep[][]{Steiner2004DetectionCarinae, Teodoro2012HePassage, Teodoro2016HeMINIMA, Davidson2015ETAFeatures}, the \ion{N}{II} lines \citep[][]{Davidson2015ETAFeatures}, H-alpha \citep[][]{Richardson2010TheEvent} and a handful of further lines (\ion{He}{I}, \ion{N}{II}, \ion{Na}{I D}, \ion{Si}{II}, \ion{Fe}{II}) during the 2009 periastron \citep[][]{Richardson2015TheEvent}.

\subsection{Radial velocities: the Balmer series}
\label{sec:radial_velocities_balmer_series}

In this study of \etacar{} we concentrate our efforts on the Balmer emission lines from H-alpha to H-kappa. These lines serve as an ideal probe into the dynamics of extended emission, thanks to their high signal-to-noise and wide coverage of different spatial scales of emission in the wind.

\subsubsection{Multi-Gaussian spectral line fitting}
\label{sec:multig_spectral_line_fitting}

To extract radial velocities from the Balmer series lines we employ multi-Gaussian fitting to decompose individual line profiles into their constituent components. This line-fitting tool has been previously applied to the complex Galactic system SS433 whose multiple modes of mass-loss are elucidated by consideration of its constituent Gaussian components \citep{Blundell2008SSMass, Blundell2011SS433sFlare}.

The approach of using multiple Gaussian components to track independent variations within complex line profiles in \etacar{} has been utilised previously \citep[eg.][]{Ruiz1984TimeCarinae, Nielsen2007Interactions}. By fitting line profiles separately we can monitor discrepancies between different line transitions. We use Gaussians as the base function because they are mathematically uncomplicated; and moreover, they often well-represent the appearance of simpler line profiles \citep{Blundell2008SSMass, Blundell2011SS433sFlare}.

For \etacar{} our multi-Gaussian fitting algorithm is implemented as follows. The continuum is removed using a Savitzky--Golay filter \citep[][]{Savitzky1964SmoothingProcedures} on regions of the spectrum not dominated by line transitions. We then construct a template, for a specific line profile, consisting of multiple Gaussian functions, each with an associated initial set of parameter guesses and bounds. The resulting model is a non-linear function constructed from the sum of the individual Gaussians. The model parameters are optimised using a Trust Region Reflective algorithm \citep[][]{Branch1999AProblems, Jones2001SciPy:Python} which efficiently explores the multidimensional parameter space in order to minimise the sum of squares.

The templates are built using a combination of human input and statistical testing. Starting from one component, we manually add new components based on visual inspection of the residuals. We repeat this process until we find a template which consistently produces good fits for spectra over a few observing nights. We allow ourselves the freedom to vary the number of components through time and use negative Gaussians to model absorption.

To evaluate the goodness of fit we follow \citet[][]{Riener2019GAUSSPY+:Spectra} by assessing whether the normalised residuals show a normal distribution, since we expect the errors in the flux bins to be Gaussian and homoscedastic. We perform a two-sided Kolmogorov--Smirnov test \citep[][]{Kolmogorov1933NoTitle, Smirnov1939OnSamples} where the null hypothesis states that the normalised residuals resemble a normal distribution. If the p-value for a multi-Gaussian fit is greater than $0.0027$ ($3\sigma$) then we fail to reject the null hypothesis and the fit is considered good. We achieve a $96\%$ success rate for the high excitation Balmer lines between H-gamma and H-kappa. For H-alpha and H-beta we are not able to satisfy the statistical test. The high signal-to-noise of these lines requires many Gaussian components to fit the observed complexity. The model-optimisation algorithm is only capable of co-fitting a finite number of free parameters without a substantial reduction in performance. Hence, for templates with greater than ${\sim} 10$ components (${\sim} 30$ free parameters) the statistical test is less instructive. We also note that H-alpha and H-beta may show further systematic effects as discussed in Section \ref{sec:colliding_winds} and \ref{sec:time_dependent_mass_loss}.

We apply our multi-Gaussian fitting algorithm to the \etacar{} spectra. Example fits are shown for apastron in Figure \ref{fig:multig_balmer_fits_apastron} and for periastron in Figure \ref{fig:multig_balmer_fits_periastron}. The individual Gaussian components (light green and orange) that make up the total fit (light blue) can be seen to vary depending on the line transition and position in the orbital cycle. H-alpha requires 14-16 components, while in contrast H-iota only requires 6-8 components, depending on phase. Despite the substantial variation we find a commonality to the templates: 4 main emission lines, a narrow low velocity component present from H-alpha to H-delta at apastron and becoming less strong near periastron, and a very broad component present from H-alpha to H-epsilon that persists across all phases.

Once the spectral lines are fit we leverage the power of multi-Gaussian fitting. By deconstructing the complex line profiles into individual components, we are able to scrutinize each component for astrophysical meaning. We note that despite using Gaussians as our base function we do not require dynamical components to be strictly Gaussian. As shown by \citet[][]{Blundell2007FluctuationsTimescales}, it is possible to take a weighted mean of multiple components (centroid wavelength of each Gaussian weighted by its area) to extract clear signals from a superposition of profile shapes.

We explore our fitting results and, at first, find no coherent signal for any of the 4 main emission components individually (orange lines in Figures \ref{fig:multig_balmer_fits_apastron} and \ref{fig:multig_balmer_fits_periastron}). However, after taking their weighted mean we extract clear radial velocity curves, as shown in Figure \ref{fig:kepler_model_fits}. Most notably, we uncover a low-amplitude radial velocity signature in H-alpha. The resulting radial velocity curves for the Balmer lines suggest that the 4 main emission components, all together, comprise the emission from \etacar{}'s primary wind.

To estimate the uncertainties we consider the velocity variance on an intra-nightly basis, assuming the velocities are not subject to real astrophysical changes on daily timescales given the orbital period of $P=2022.7 \days{}$. The final uncertainties for our radial velocity data are then calculated by the pooled variance, $s_{\scriptscriptstyle \rm{RV}}^{2}$, of intra-nightly spectra, defined as
\begin{equation}
s_{\scriptscriptstyle \rm{RV}}^{2} = \frac{\sum_{i=1}^{k} (n_{i} - 1)s_{i}^{2}}{\sum_{i=1}^{k} (n_{i} - 1)},
	\label{eq:pooled_variance}
\end{equation}
where $n_{i}$ is the number of observations in a given night and $s_{i}$ is the intra-nightly sample variance. We could allow the uncertainties to vary in time, but due to the number of observations per night being low we can obtain a better overall estimate of the uncertainties by combining the intra-nightly variance across all times.

\subsubsection{A benchmark algorithm for radial velocity extraction}
\label{sec:benchmark_algorithm_for_rv_extraction}

There exist a handful of alternative techniques for extracting radial velocities in mainstream usage. For computing the bulk velocity of emission lines there are basic algorithms, such as finding the emission-line peak or the bisector at half the maximum flux (BHM), or more sophisticated algorithms, such as cross-correlation. Additionally, when extracting radial velocities from wind lines displaying P Cygni profiles the absorption minimum is often used as a proxy for the stellar motion.

To verify the radial velocities output from our multi-Gaussian fitting algorithm, we implement the BHM algorithm on \etacar{}'s Balmer lines. The BHM output will serve as a simple benchmark to compare our results against.

\section{Modelling}
\label{sec:modelling}

We model the dynamics of the Balmer lines' emission (H-alpha to H-kappa) originating from the primary wind of \etacar{} based on the radial velocities computed in Section \ref{sec:observations}. For all models we assume an orbital period of $P=2022.7 \days{}$ \citep[][]{Damineli2008TheEvents}.

\subsection{Keplerian motion}
\label{sec:keplerian_motion}

\begin{figure*}
	\includegraphics[width=\textwidth]{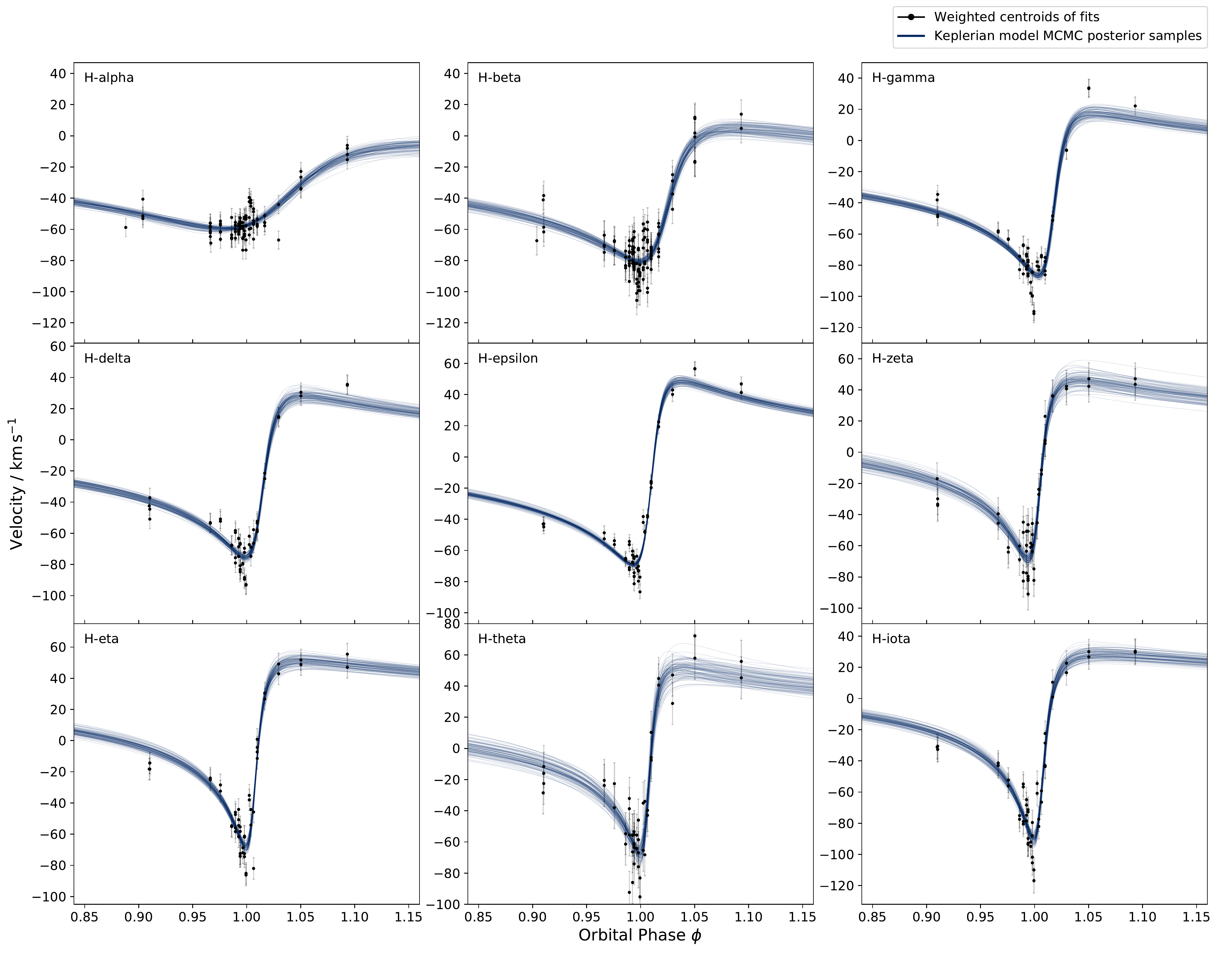}
    \caption{Keplerian models independently fit to the radial velocity observations of \etacar{}'s Balmer series. Each panel shows the Balmer line, 100 samples from the converged MCMC run (dark blue) and the radial velocities (black) extracted from the multi-Gaussian algorithm and associated uncertainties. The radial velocities are the weighted mean of the 4 main emission components (orange Gaussians) as shown in Figures \ref{fig:multig_balmer_fits_apastron} and \ref{fig:multig_balmer_fits_periastron}.}
    \label{fig:kepler_model_fits}
\end{figure*}

As an instructive exercise, we start by assuming the entire line emission region tracks the orbital motion of the primary star. If this is the case the radial velocities will correspond to Keplerian orbits projected onto the line-of-sight. Using the standard formalism for a binary system \citep[][]{Aitken1964TheStars} this is written
\begin{equation}
v_{\rm{kep}}(t) = k(\cos (\omega + \nu) + e\cos{\omega}) + \gamma,
	\label{eq:keplerian_los_velcoity}
\end{equation}
where $v_{\rm{kep}}$ is the Keplerian velocity projected onto the line-of-sight, $k$ is the radial velocity semi-amplitude, $e$ is the eccentricity, $\omega$ is the longitude of periastron, $\nu$ is the true anomaly and $\gamma$ is the radial velocity offset. The true anomaly is calculated from the orbital phase, $\phi$, via the mean anomaly, $M$, and eccentric anomaly, $E$, related by
\begin{align}
    \label{eq:phase_and_mean_anomaly}
& 2 \pi \phi = M = E - e \sin E, \\
& \tan \Big( \frac{\nu}{2} \Big) = \sqrt{\frac{1 + e}{1 -e}} \tan \Big( \frac{E}{2} \Big).
    \label{eq:phase_mean_eccentric_anomaly}
\end{align}
We solve Equation \ref{eq:phase_and_mean_anomaly}, known as Kepler's equation, iteratively using the Newton-Raphson root-finding algorithm. The phase is calculated from the period and time of periastron, $T_{0}$.

We fit the radial velocity observations of \etacar{}'s Balmer lines with the Keplerian model using the affine-invariant Markov Chain Monte Carlo (MCMC) algorithm, \texttt{emcee}, implemented by \citet[][]{Foreman-Mackey2013Emcee:Hammer}. We allow the parameters in the set $\theta = \{T_{0}, e, k, \omega, \gamma \}$ to vary. The log-likelihood function, applied to each line independently, for the Keplerian model, $v_{\rm{kep}}$, compared to the radial velocity data, $D_{\phi}$, and associated uncertainties, $\sigma_{\phi}$, is
\begin{equation}
\ln P(\theta \mid D, \sigma) = -\frac{1}{2} \sum_{\phi} \Bigg\lbrack \frac{(D_{\phi} - v_{\rm{kep}}(\theta)_{\phi})^{2}}{{\sigma_{\phi}}^{2}} + \ln (2 \pi {\sigma_{\phi}}^{2}) \Bigg\rbrack.
	\label{eq:log_liklihood_kepler_mcmc}
\end{equation}
The algorithm is run with 128 walkers for 6000 burn in steps and then for a further 6000 sampling steps. In our testing we find chains of this length result in good convergence (using the autocorrelation time \citep[][]{Foreman-Mackey2013Emcee:Hammer} as a guiding metric) and well-resolved parameter posteriors.

\subsubsection{Keplerian Results}
\label{sec:keplerian_results}

\begin{figure}
	\includegraphics[width=1\columnwidth]{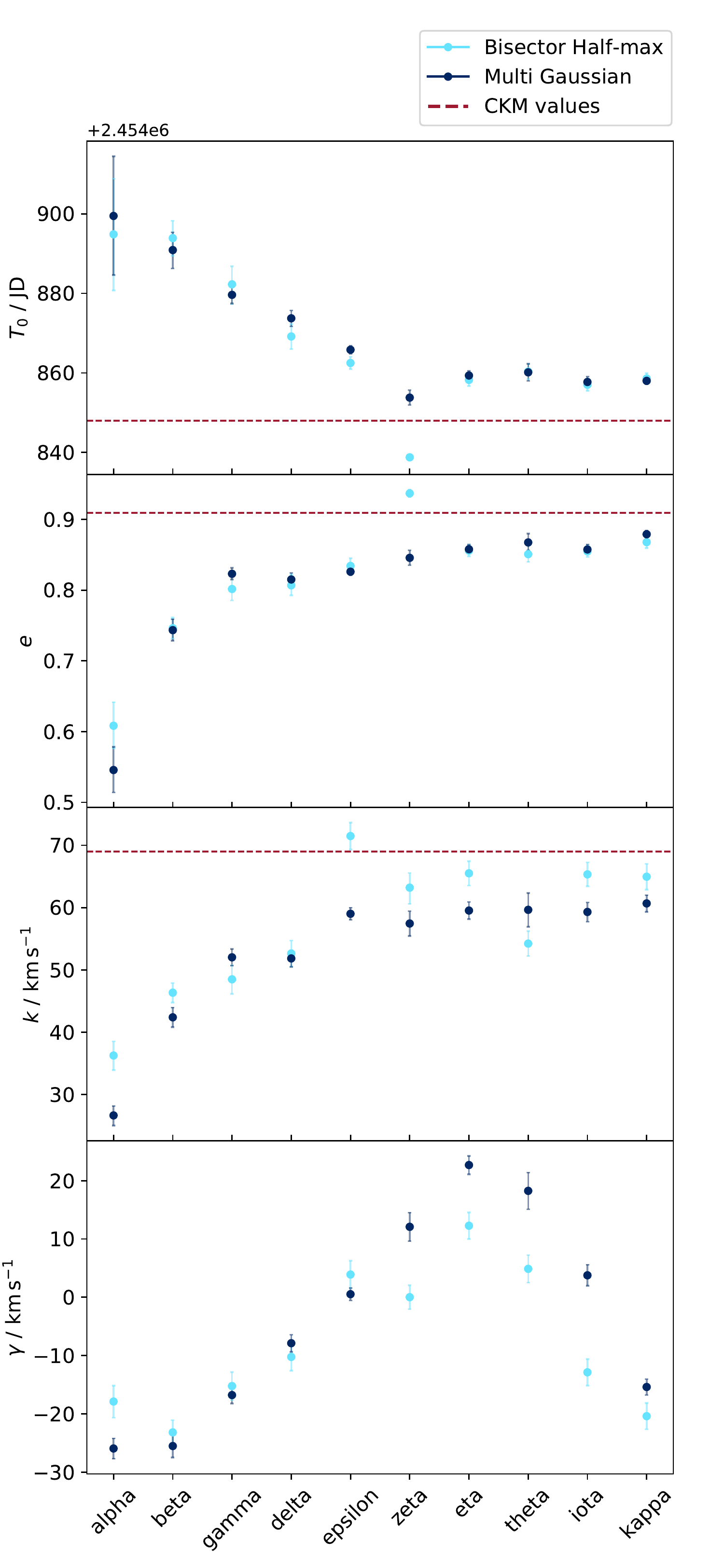}
    \caption{Trends in the parameters output from Keplerian models applied to the multi-Gaussian (dark blue) and BHM (light blue) radial velocity observations of \etacar{}. The final convolutional Keplerian motion (CKM) model parameters are shown as the red dashed asymptotes (see Section \ref{sec:ck_motion} for details).}
    \label{fig:kepler_model_trends}
\end{figure}

In Figure \ref{fig:kepler_model_fits} we show the results of applying the Keplerian model independently to a range of different Balmer lines' radial velocities. Each panel corresponds to a different model run, where the Balmer line is labelled in the top left. We plot $v_{\rm{kep}}$ for 100 random samples from the MCMC run against the radial velocity observations extracted using the multi-Gaussian algorithm. The radial velocities are the weighted mean of the 4 main emission components (orange Gaussians) as shown in Figures \ref{fig:multig_balmer_fits_apastron} and \ref{fig:multig_balmer_fits_periastron}. The Keplerian model is capable of producing good fits to all of the Balmer series lines. However, as expected the parameters derived using this model do not agree between lines. The dependence of the Keplerian radial velocity curves on $e$ and $\omega$ is shown in Appendix \ref{sec:keplerian_radial_velocity_curves}.

In Figure \ref{fig:kepler_model_trends} we plot the optimised parameters from all of the MCMC runs for both the multi-Gaussian (dark blue) and BHM (light blue) methods of radial velocity extraction. We show the parameters' median values and 16th-84th percentile uncertainties from their posterior distributions. The parameters show correlations across the Balmer series when ordered by transition energy. We find very clear monotonic trends for $T_{0}$, $e$ and $k$ -- whilst $\gamma$ exhibits more complex behaviour. The trends are less noisy in the multi-Gaussian results relative to the BHM results indicating the superiority of this extraction technique. The trends show that the higher energy Balmer lines have associated radial velocity curves that appear phase shifted later in time, as well as less eccentric and lower in amplitude. These effects have also been observed between different species in \etacar{}. \citet[][]{Nielsen2007Interactions} noted the radial velocities of their Balmer lines appeared offset in phase and reduced in amplitude relative to He I, and \citet[][]{Davidson2000} found their higher series Paschen lines showed larger velocity shifts. The monotonic trends are also asymptotic: the higher energy lines appear to converge towards a single set of parameter values.

We suggest the systematic sequencing of these results is a consequence of probing different emission-line regions for each Balmer line. The higher excitation lines are preferentially emitted in denser and hotter regions relative to lower excitation Balmer lines. Hence, lines lower down the Balmer series are emitted in progressively more extended regions further away from the star, with H-alpha being emitted over the largest and most extended volume. For extended line emission the radial velocity measurements will contain dynamical information which is not purely Keplerian. The more extended the line emission, the greater the departure from the true Keplerian motion the observations will imply. Conversely, lines towards the top of the Balmer series, approaching the asymptote in Figure \ref{fig:kepler_model_trends}, are formed at increasingly close regions to the central star, deeper in the wind, and as such the parameters become increasingly representative of the true orbital motion.

\subsection{Convolutional Keplerian motion}
\label{sec:ck_motion}

We formulate a semi-analytical model to account for the dynamical information encoded in lines formed in extended regions. The model is a first attempt to encapsulate both the star's orbital motion, and the propagation of the wind by accounting for its time delay to the location where the photons are actually emitted.

To explain the net motion observed from the volume of gas in which an emission line is formed we simplify the picture of a binary star with a powerful wind to the following paradigm:
\begin{itemize}
  \item Gas in the stellar wind at radii less than the sonic point co-orbits with the star. Emission from these regions will appear Keplerian.
  \item Outside of the sonic point there is no way for gas to respond to the acceleration of the central star via pressure waves. The gas moves outwards, and is subject to radiative acceleration.
  \item At each radius in the stellar wind the radial velocities of an infinitesimally thin shell average to zero, assuming a steady spherically symmetric wind. However, the shell has an overall velocity equal to the orbital velocity of the star at the time when it was accelerated past the sonic point.
  \item Therefore, the total stellar wind can be thought to consist of shells, each encoding the Keplerian motion of the star from a time in the past defined by the flow time of the wind from where the sonic point was previously to the location where the photons are emitted.
  \item As such, the net motion of the total line emission region is an average of the Keplerian velocities over an interval backwards in time which equates to the flow time of the wind to reach the emission radii. In effect, this is a moving average of the Keplerian velocities.
  \item Further to this, depending on the line transition, lines are emitted at various strengths over a range of different radii, and thus different flow times. So, the moving average should be weighted by the luminosity of the line transition for a given species at a given radius.
\end{itemize}

This description is idealised in order to assess its effectiveness, which we explore in Sections \ref{sec:timescales} and \ref{sec:outflows_eta_car}. Using this framework we can now define our model. More formally, a weighted moving average of the Keplerian velocities can be be written as a convolution 
\begin{equation}
v_{\rm{ckm,n}}(t) = v_{\rm{kep}}(t) * \lambda_{L,n}(t) = \int_{-\infty}^{\infty} v_{\rm{kep}}(\tau) \lambda_{L,n}(t - \tau) d\tau,
	\label{eq:convolutional_keplerian_motion}
\end{equation}
where $v_{\rm{ckm,n}}$ is the convolutional Keplerian motion (CKM), $v_{\rm{kep}}$ is the Keplerian velocity from Equation \ref{eq:keplerian_los_velcoity} and $\lambda_{L,n}(t)$ is the line-formation kernel for each emission line, $n$, which weights $v_{\rm{kep}}$ as a function of wind flow time. 

An alternative way to think about CKM is by drawing parallels with the retarded potential in electrodynamics \citep[][]{Griffiths2007IntroductionElectrodynamics}. The stellar wind's velocity field generated by the orbital motion is similar in principle to the electromagnetic field generated by a time-varying charge distribution. Any changes to the central source take a finite time to propagate outwards -- in our case at the flow speed of the wind, and in electrodynamics at the speed of light.

\subsubsection{Radiative transfer and the line-formation kernels}
\label{sec:radiative_transfer}

In order to compute the line-formation kernels, $\lambda_{L,n}(t)$, we require knowledge of a specific line's emission strength as a function of radius in the stellar wind. This quantity cannot be calculated analytically because it depends on solving a coupled problem: the radiation field depends on the emission and extinction of the medium which themselves depend on the radiation field. Accordingly, we employ the non-local thermodynamic equilibrium, fully line-blanketed stellar atmosphere code, \texttt{CMFGEN} \citep[][]{Hillier1998TheOutflows, Hillier2001CMFGEN}. \texttt{CMFGEN} solves the radiative transfer equation in the comoving frame, subject to the constraints of radiative and statistical equilibrium. We select \texttt{CMFGEN} because of its historical use in modelling \etacar{}, making our results more comparable to past work. In practice, any stellar atmosphere code could be used and the results of our CKM model should be independent of this choice.

We use \texttt{CMFGEN} to find the linear luminosity density (otherwise known as luminosity per unit radius) as a function of radius, $\lambda_{L,n}(r)$, for each line transition observed in \etacar{}. The line-formation kernels are then calculated by converting $\lambda_{L,n}(r)$ to be a normalised function of wind flow time by integrating over the velocity profile of the wind using the following equations
\begin{align}
    \label{eq:beta_velocity_law}
& v_{\rm{wind}}(r) = v_{0} + (v_{\infty} - v_{0}) \Big(1 - \frac{R_{*}}{r} \Big)^{\beta}, \\
	\label{eq:integrate_velocity_for_time}
& t_{\rm{flow}} = \int_{r_{s}}^{r} \frac{dr}{v_{\rm{wind}}(r)},
\end{align}
where $v_{0}$ is the initial wind velocity, $v_{\infty}$ is the terminal wind velocity, $R_{*}$ is the stellar radius, $r_{s}$ is the sonic radius and $\beta$ is a free parameter describing the steepness of the velocity law. In Equation \ref{eq:beta_velocity_law} we assume that the wind is accelerated by a standard $\beta$-law as first derived by \citet[][]{Castor1975Radiation-drivenStars} for line-driven winds and we set $\beta=1$.

For \etacar{}, and most stars with powerful winds, this is likely a good approximate form for the conversion. We note, however, it has been suggested that \etacar{}'s wind may be super-Eddington and its velocity profile would therefore be a modified $\beta$-law \citep[][]{Owocki2004ALimit}. However, in general the line-formation region is extended well beyond the acceleration region of the wind and so the results will be largely insensitive to the velocity prescription.

The main dependence of the $\lambda_{L,n}(r)$ curves is on a few stellar parameters input into the code. In principle, it is possible to create a multidimensional grid of $\lambda_{L,n}(r)$ curves, spanning a few key parameters, and then interpolate between the results by including the parameters in our CKM modelling routine. Thereby, it would be possible to match radial velocities and stellar types in one unique solution. Of particular interest are the stellar radius, $R_{*}$, mass-loss rate, $\Dot{M}$, and the terminal wind velocity, $v_{\infty}$. Unfortunately, for \etacar{} the effect of $R_{*}$ has a negligible effect on the output. As previously found by \citet[][]{Hillier2001} the extreme mass loss prevents the determination of $R_{*}$ accurately. We test the effect of changing the inner boundary of the simulation on $\lambda_{L,n}(r)$ and reach the same conclusion: the results are insensitive to changes in $R_{*}$ and thus we cannot calculate an accurate effective temperature for \etacar{} (see Appendix \ref{sec:inner_boundary_testing}). Likewise, we cannot determine values for $\Dot{M}$ and $v_{\infty}$ because $\lambda_{L,n}(r)$ is degenerate for these two parameters (see Appendix \ref{sec:line_formation_kernel_degeneracy}). 

\begin{figure}
	\includegraphics[width=1\columnwidth]{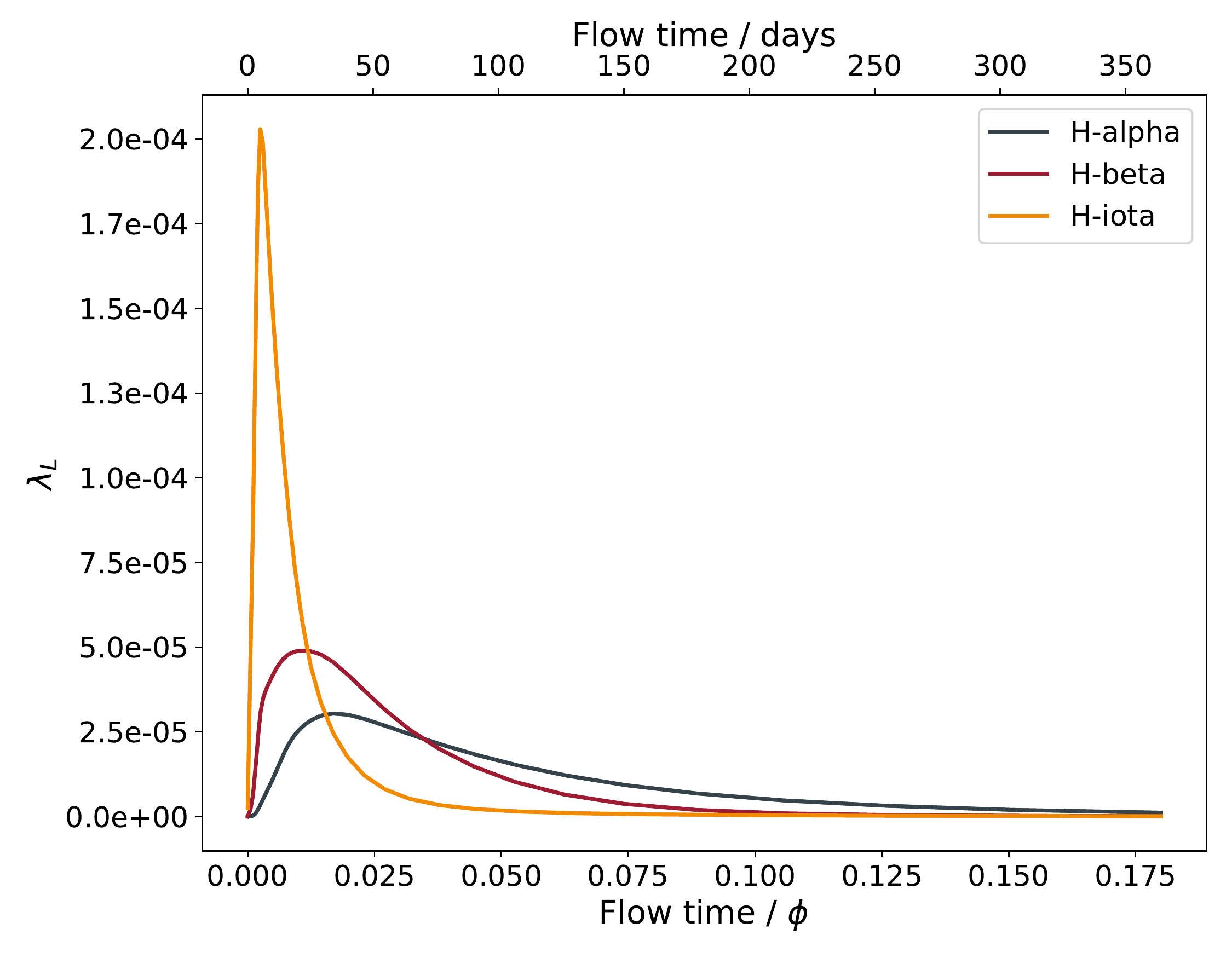}
    \caption{Line-formation kernels for an example set of 3 Balmer lines for \etacar{}. Each kernel is the linear luminosity density (otherwise known as luminosity per unit radius) as a function of wind flow time beyond the sonic point. Flow time is shown in days (top x axis) and in units of \etacar{}'s orbital phase (bottom x axis).}
    \label{fig:ckm_kernels}
\end{figure}

\begin{figure*}
	\includegraphics[width=2\columnwidth]{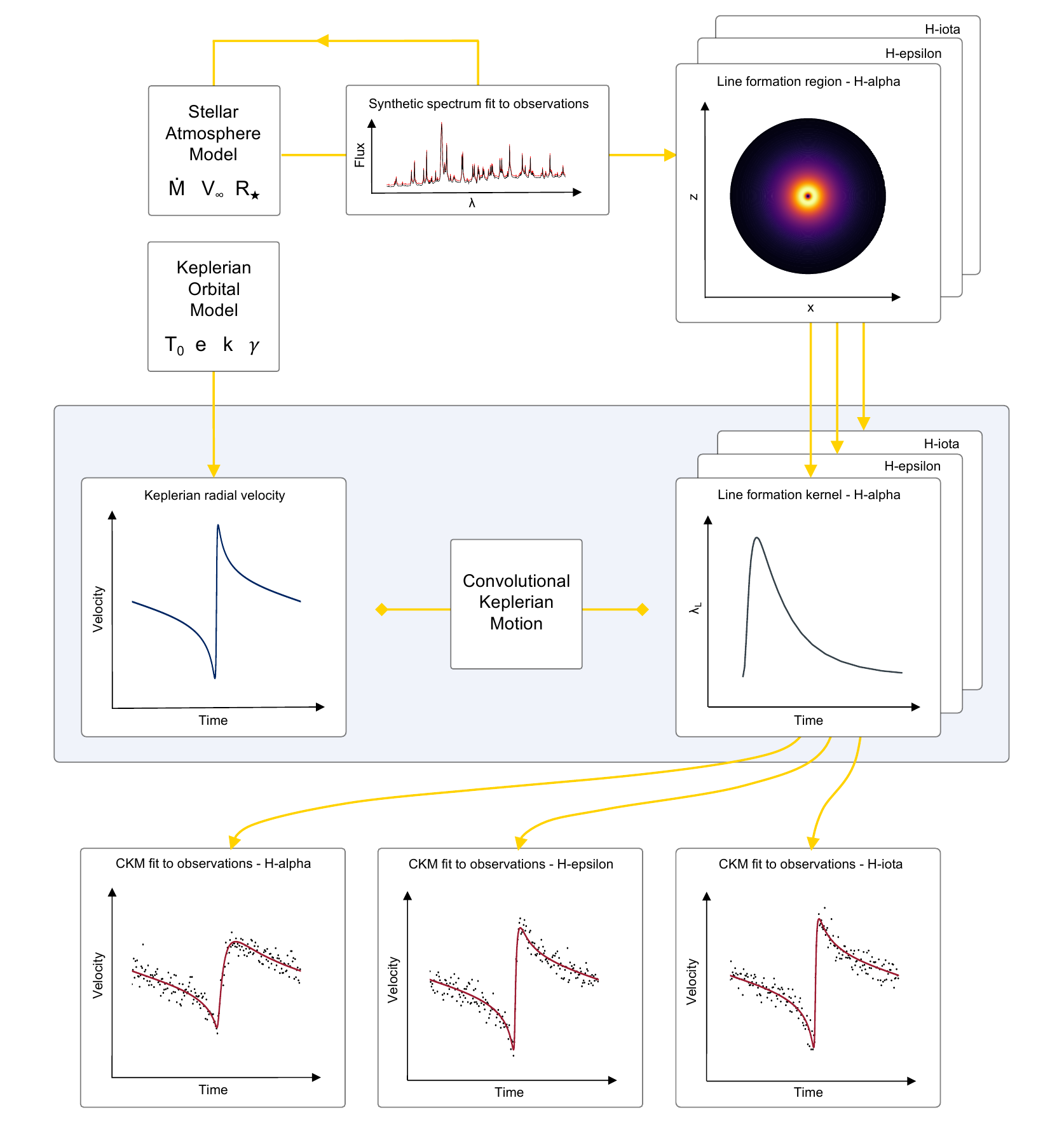}
    \caption{CKM model schematic for an example set of 3 Balmer lines. The stellar atmosphere model iteratively fits a synthetic spectrum to an observed spectrum to constrain the key stellar parameters: mass-loss rate, $\Dot{M}$, terminal wind velocity, $v_{\infty}$ and stellar radius, $R_{*}$. The best fit stellar atmosphere model outputs a set of line-formation regions which are converted to 1D kernels as a function of wind flow time. The Keplerian orbital model is convolved with each line-formation kernel (shaded box) to produce a set of CKM models to compare to radial velocity observations. By optimising the Keplerian input parameters -- time of periastron, $T_{0}$, radial velocity semi-amplitude, $k$, eccentricity, $e$, longitude of periastron, $\omega$ -- to best fit the observations we can infer the true orbital motion from extended emission.}
    \label{fig:ckm_schematic}
\end{figure*}

\begin{figure*}
	\includegraphics[width=2\columnwidth]{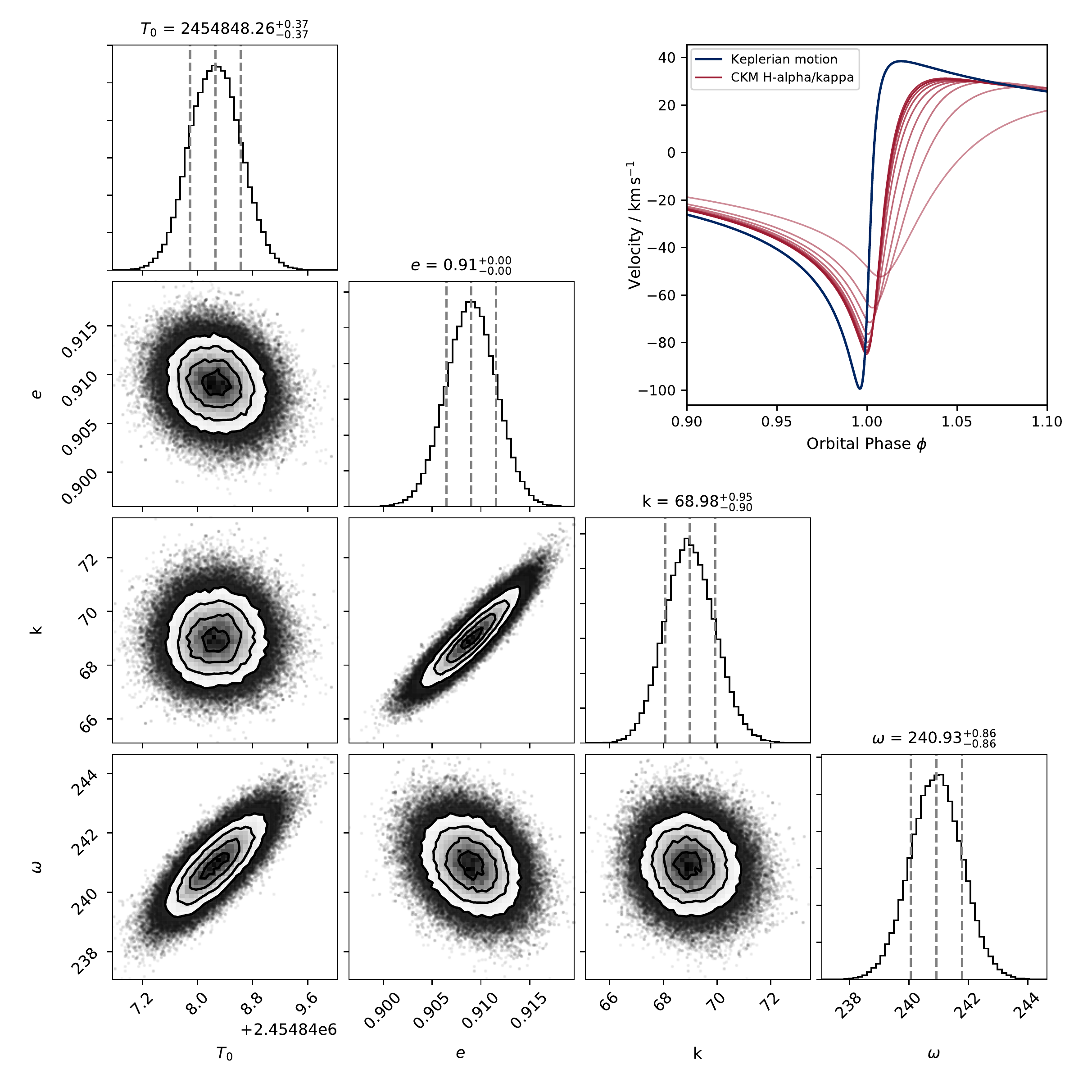}
    \caption{Bottom left corner: posterior probability distributions of the CKM model's free parameters: time of periastron, $T_{0}$, in Julian days; eccentricity, $e$; radial velocity semi-amplitude, $k$, in $\rm{\,km \, s^{-1}}$; and longitude of periastron, $\omega$, in degrees. In the 1D histograms the dotted lines indicate the 16th, 50th and 84th percentiles. In the 2D histograms the contours represent the $1\sigma$, $2\sigma$ and $3\sigma$ statistical uncertainties. Top right panel: set of CKM model curves (red) for the Balmer lines, H-alpha to H-kappa (without $\gamma$ correction), and the inferred Keplerian motion (dark blue) of the best fit model.}
    \label{fig:ckm_corner}
\end{figure*}

\begin{figure*}
	\includegraphics[width=\textwidth]{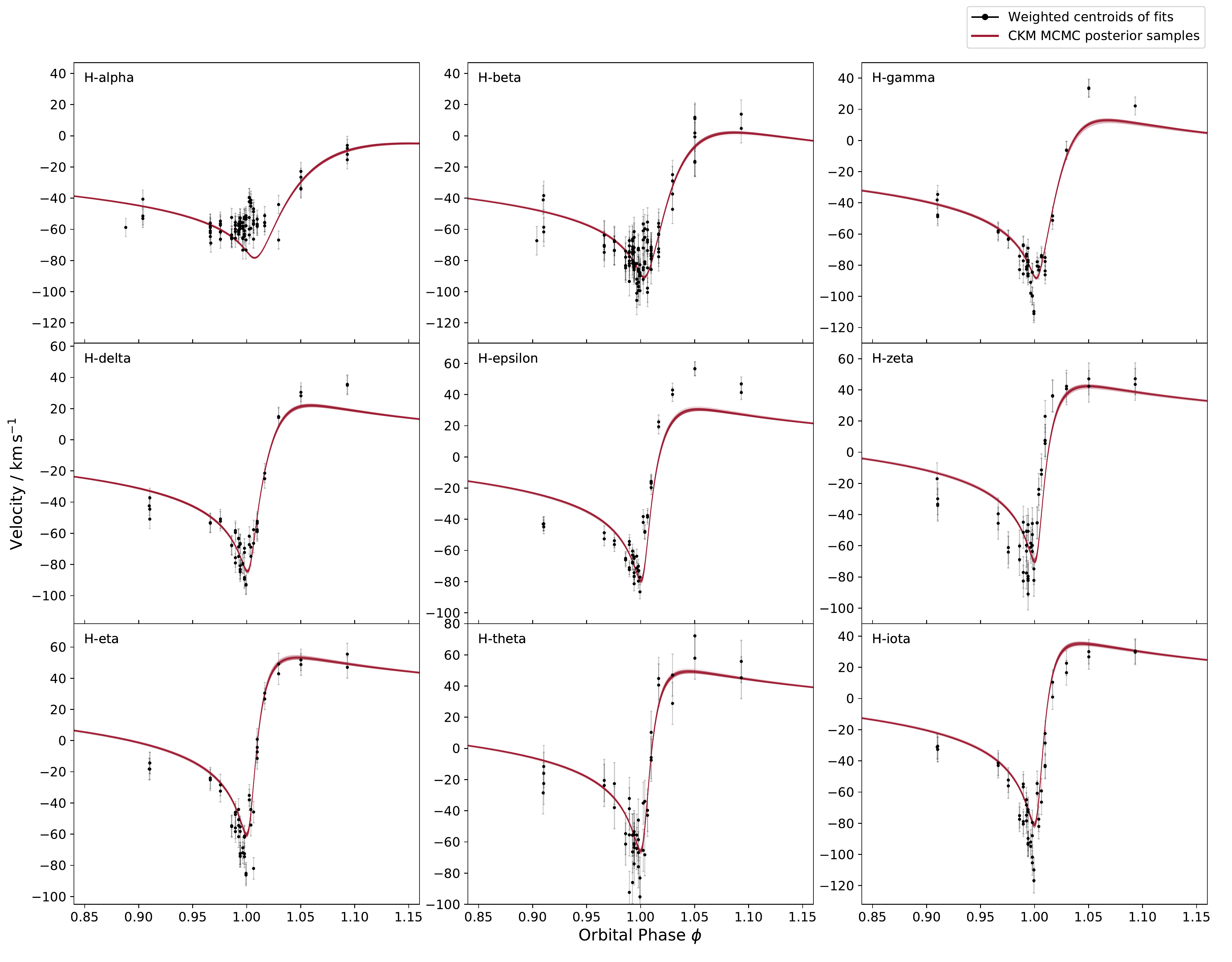}
    \caption{CKM model co-fit to the radial velocity observations of \etacar{}'s Balmer series. Each panel shows the Balmer line, 100 samples from the converged MCMC run (red) and the radial velocities (black) extracted from the multi-Gaussian algorithm and associated uncertainties. The radial velocities are the weighted mean of the 4 main emission components (orange Gaussians) as shown in Figures \ref{fig:multig_balmer_fits_apastron} and \ref{fig:multig_balmer_fits_periastron}.}
    \label{fig:ckm_model_fits}
\end{figure*}

To overcome the difficulties in determining the stellar parameters we introduce a preliminary step into the modelling routine. Rather than only utilising \texttt{CMFGEN}'s output line-formation regions, we first constrain the stellar parameters by ensuring a good fit between the synthetic spectrum and the observations. In this way we are now incorporating the information from the total observed spectrum, to lock down $R_{*}$, $\Dot{M}$ and $v_{\infty}$, as well as the radial velocities extracted from a time series of spectra. By forcing our model to fit both a spectral snapshot and the time-domain motion, we constrain the results more.

The final wind parameter that has a significant effect on $\lambda_{L,n}(r)$ is the clumping. A standard line-driven wind is expected to have strong clumping that develops from the intrinsic instability in line driving \citep[][]{MacGregor1979RadiativeStars, Owocki1988Time-dependentModel}, and this effect has been observationally substantiated \citep[eg.][]{Ebbets1982TheSupergiants, Kaper1996Long-Stars, Oskinova2006High-resolutionStars}. A clumpy wind creates higher densities at larger radii; therefore, line emission is possible in more extended regions from the star than previously allowed in a homogeneous wind (see Appendix \ref{sec:clumping_testing}). For \etacar{} we use the same parameterisation as \citet[][]{Hillier2001} and \citet[][]{Groh2012OnSpectra}: a volume filling factor, $f_{\rm{v}}=0.1$, which describes the clumping at infinity and $v_{\rm{cl}}=100 \kmpers{}$ which denotes the wind velocity at which clumping becomes important. We note, however, that owing to the extremely high mass-loss rate in \etacar{}, the line-driving instability may not be a strong factor. But, if the wind is instead continuum driven then it may be regulated by ``porosity", which reduces the effective opacity in deep layers \citep[][]{Owocki2004ALimit}. In either case, we expect over-densities to be present in the wind and we have chosen standard parameter values to account for this effect in the radiative transfer simulation.

For \etacar{} we fix $R_{*}=60 \solarm{}$ and for $\Dot{M}$ and $v_{\infty}$ we use the values found by \citet[][also using \texttt{CMFGEN}]{Groh2012OnSpectra}, of $\Dot{M}=8.5 \times 10^{-4} \masslossrate{}$ and $v_{\infty}=420 \kmpers{}$. We show the resulting line-formation kernels for H-alpha, H-beta and H-iota in Figure \ref{fig:ckm_kernels}.

\subsubsection{Modelling routine}
\label{sec:ckm_modelling_routine}

In Figure \ref{fig:ckm_schematic} we present a schematic of the CKM modelling routine. The schematic shows 3 example lines (H-alpha, H-epsilon, H-iota) to illustrate the difference in output between lines formed in different regions of the wind, but the model can be applied to any emission line or set of lines for any emission-line star. As described in Section \ref{sec:radiative_transfer}, a stellar atmosphere code is employed to fit a synthetic spectrum to observations of a chosen target. By iteratively comparing the stellar atmosphere model to the observations we identify the best fitting values for the key stellar parameters such as $R_{*}$, $\Dot{M}$ and $v_{\infty}$. Using the best fit model, a set of $\lambda_{L,n}(r)$ curves are generated: one for each emission line for which we have corresponding radial velocity observations. At this stage we are modelling the wind with no latitudinal dependence and so the line-formation regions are spherically symmetric. $\lambda_{L,n}(r)$ is converted to a function of wind flow time, $\lambda_{L,n}(t)$, using Equations \ref{eq:beta_velocity_law} and \ref{eq:integrate_velocity_for_time}, and summed over all directions producing a 1D line-formation kernel for each line. A Keplerian orbital model is created using Equation \ref{eq:keplerian_los_velcoity} and convolved with each line-formation kernel (shaded box in Figure \ref{fig:ckm_schematic}). The resulting set of CKM curves are compared to radial velocity observations.

For our case study of \etacar{} we fit the CKM curves to the radial velocities extracted from the Balmer lines -- H-alpha to H-kappa. As described in Section \ref{sec:keplerian_motion}, we using the affine-invariant MCMC algorithm, \texttt{emcee}, implemented by \citet[][]{Foreman-Mackey2013Emcee:Hammer} to optimise the model fitting. We allow the parameters in the set $\theta = \{T_{0}, e, k, \omega \}$ to vary. We fix $\gamma$ for each line to be the same as the results in Figure \ref{fig:kepler_model_trends}. By using the $\gamma$ values calculated from individually optimising the Keplerian model for each line, we allow the flexibility for $\gamma$ to be different between lines, whilst reducing the number of free parameters by the number of lines in the radial velocity set. The log-likelihood function for our CKM model, co-fit to all the Balmer lines concurrently, is equal to the sum of log-likelihoods for each line:
\begin{align}
& \ln P(\theta \mid D, \sigma) = \sum_{n} \ln P(\theta \mid D_{n}, \sigma_{n})
	\label{eq:log_liklihood_total_ckm_mcmc} \\
& = -\frac{1}{2} \sum_{n} \sum_{\phi} \Bigg\lbrack \frac{(D_{\phi,n} - v_{\rm{ckm},n}(\theta)_{\phi})^{2}}{{\sigma_{\phi,n}}^{2}} + \ln (2 \pi {\sigma_{\phi,n}}^{2}) \Bigg\rbrack,
    \label{eq:log_liklihood_per_line_ckm_mcmc}
\end{align}
where $v_{\rm{ckm},n}$, $D_{\phi,n}$ and $\sigma_{\phi,n}$ are the CKM model, radial velocity data and associated uncertainties for a given Balmer line, $n$, respectively. The algorithm is run with 128 walkers for 6000 burn in steps and then for a further 6000 sampling steps to ensure convergence and well-resolved parameter posteriors.

\subsubsection{CKM Results}
\label{sec:ckm_results}

In Figure \ref{fig:ckm_corner} we show the posterior probability distributions for each of the parameters in the set $\theta$ from the MCMC run. The model converges to a unique solution with the 1D histograms showing approximate Gaussian distributions for all of the free parameters. The best fit values are $T_{0}=2454848$ (JD), $e=0.91$, $k=69 \kmpers{}$ and $\omega=241^\circ$, with small statistical uncertainties on each value. The optimised set of CKM curves (red) can be seen in the top right panel, for lines H-alpha to H-kappa, along with the inferred Keplerian motion (dark blue) of the central star. The CKM velocity profile for H-alpha shows the greatest departure from the Keplerian motion; it is phase shifted later in time, less eccentric and lower in amplitude owing to it having the most extended line-formation kernel. The rest of the Balmer lines show CKM curves becoming asymptotically more similar to the underlying Keplerian motion as the transition energy increases. These results qualitatively emulate the parameter variations found between different lines, when assuming Keplerian motion, as seen previously in Figure \ref{fig:kepler_model_trends}.

In Figure \ref{fig:ckm_model_fits} we compare the best fit CKM model against the radial velocity observations extracted using the multi-Gaussian algorithm. In each panel, we plot $v_{n,\rm{ckm}}$ for 100 random samples from the MCMC run for a Balmer line, $n$, labelled in the top left. Overall, the model is able to reproduce the progression in Balmer line radial velocity profiles. The fits are particularly good at times close to periastron for Balmer lines between H-gamma and H-iota. The model struggles to fit the profiles in the wake of periastron, around $\phi {\sim} 1.05$, for a few lines. Further to this, the model is not able to fit the extent to which the velocities are modified in H-alpha, and to some extent H-beta. However, these remaining deviations may be caused by external factors we have not included in the model; in particular, the influence of the companion which will predominantly affect the lines formed in the most extended regions (see Section \ref{sec:outflows_eta_car} for further discussion).

\begin{figure}
	\includegraphics[width=1\columnwidth]{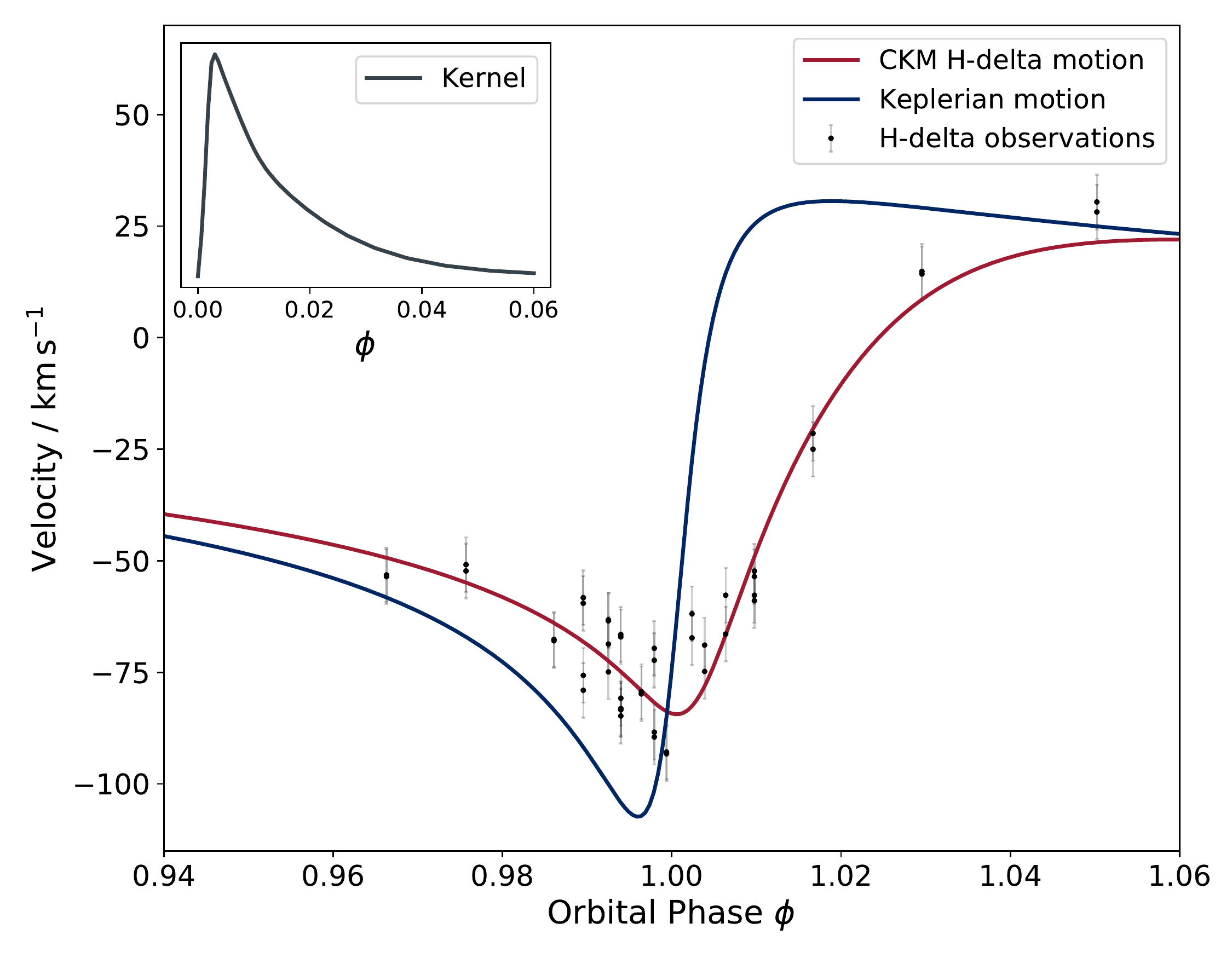}
    \caption{CKM best fitting model (red) to the H-delta radial velocities (black) extracted from the multi-Gaussian algorithm and associated uncertainties. Also shown is the inferred Keplerian motion (dark blue) of \etacar{} and the H-delta line-formation kernel (top left inset panel) as a function of orbital phase.}
    \label{fig:ckm_parts_hdelta}
\end{figure}

In Figure \ref{fig:ckm_parts_hdelta} we focus on H-delta to show in detail the constituent parts of the CKM model.
H-delta shows line formation with wind flow times up to $6\%$ of an orbital period. As a result, when the Keplerian motion (dark blue) is convolved with this line-formation kernel (top left inset panel) the resulting CKM motion is modified significantly. In this case we see the CKM curve produces a good fit to the observations. This plot showcases the capability of our model to uncover significant differences between the underlying orbital motion and the observed motion.

\subsubsection{CKM cross-validation}
\label{sec:ckm_cross_validation}

\begin{table}
\scriptsize
  \begin{tabular}{l @{\hspace{-.2\tabcolsep}} c @{\hspace{.2\tabcolsep}} c c c c}
  \hline
  Line(s) & $T_{0}$ & $e$ & $k$ & $\omega$ & $\gamma$\\
  & (JD $+2454800$) & & ($\rm{\,km \, s^{-1}}$) & (deg) & ($\rm{\,km \, s^{-1}}$)\\
  \hline
  H-alpha & $27.4^{+16.7}_{-15.0}$ & $0.63^{+0.03}_{-0.03}$ & $31.7^{+1.1}_{-0.1}$ & $247^{+7}_{-7}$ & $-25.9^{+1.8}_{-1.7}$\\
  \\
  H-beta & $54.9^{+4.5}_{-4.1}$ & $0.82^{+0.02}_{-0.02}$ & $53.0^{+2.1}_{-1.9}$ & $254^{+4}_{-4}$ & $-25.5^{+2.0}_{-2.0}$\\
  \\
  H-gamma & $53.8^{+2.4}_{-2.3}$ & $0.90^{+0.01}_{-0.01}$ & $73.3^{+3.4}_{-2.9}$ & $252^{+4}_{-4}$ & $-16.8^{+1.4}_{-1.4}$\\
  \\
  H-delta & $54.1^{+1.8}_{-1.8}$ & $0.87^{+0.01}_{-0.01}$ & $64.6^{+2.1}_{-2.0}$ & $254^{+3}_{-3}$ & $-7.9^{+1.5}_{-1.4}$\\
  \\
  H-epsilon & $52.4^{+0.3}_{-0.5}$ & $0.87^{+0.01}_{-0.01}$ & $72.5^{+1.3}_{-1.3}$ & $267^{+1}_{-1}$ & $0.5^{+1.1}_{-1.1}$\\
  \\
  H-zeta & $41.5^{+1.9}_{-1.8}$ & $0.89^{+0.01}_{-0.01}$ & $70.5^{+2.8}_{-2.6}$ & $249^{+3}_{-4}$ & $12.1^{+2.5}_{-2.4}$\\
  \\
  H-eta & $45.7^{+1.6}_{-1.0}$ & $0.89^{+0.01}_{-0.01}$ & $73.0^{+2.1}_{-2.0}$ & $235^{+3}_{-2}$ & $22.7^{+1.6}_{-1.6}$\\
  \\
  H-theta & $48.2^{+2.2}_{-2.0}$ & $0.90^{+0.01}_{-0.01}$ & $73.9^{+5.4}_{-4.3}$ & $244^{+5}_{-5}$ & $18.3^{+3.1}_{-3.1}$\\
  \\
  H-iota & $45.2^{+1.2}_{-0.9}$ & $0.90^{+0.01}_{-0.01}$ & $72.6^{+2.6}_{-2.4}$ & $229^{+3}_{-2}$ & $3.8^{+1.8}_{-1.8}$\\
  \\
  H-kappa & $46.9^{+0.8}_{-0.7}$ & $0.92^{+0.01}_{-0.01}$ & $78.8^{+3.5}_{-3.1}$ & $228^{+2}_{-2}$ & $-15.4^{+1.4}_{-1.3}$\\
  \hline
  H-alpha/kappa & $48.3^{+0.4}_{-0.4}$ & $0.91^{+0.00}_{-0.00}$ & $69.0^{+0.9}_{-0.9}$ & $241^{+1}_{-1}$ & $*$\\
  \\
  H-gamma/kappa & $48.4^{+0.4}_{-0.4}$ & $0.89^{+0.00}_{-0.00}$ & $69.9^{+0.8}_{-0.8}$ & $246^{+1}_{-1}$ & $*$\\
  \hline
  \end{tabular}
  \caption{Top section: cross-validation of CKM fits to individual Balmer lines in \etacar{}. Bottom section: results from co-fitting to a set of Balmer lines in the interval H-alpha to H-kappa and H-gamma to H-kappa. Parameters: time of periastron, $T_{0}$, eccentricity, $e$, radial velocity semi-amplitude, $k$, longitude of periastron, $\omega$, and systemic velocity, $\gamma$. The systemic velocities used in the fits to multiple lines ($*$) are the same as in the top section for each line (see text for details). Uncertainties are purely statistical based on the model fitting.}
    \label{tab:ckm_cross_validation}
\end{table}

After fitting our CKM model to a set of \etacar{}'s Balmer lines, it serves as an interesting test to fit each line independently and see if the model recovers similar orbital parameters. In the top section of Table \ref{tab:ckm_cross_validation} we present the results of this cross-validation test. For comparison, the parameters from the original run are shown in the bottom section of the table labelled as H-alpha/kappa.  Note that the uncertainties are purely statistical and result from the model fitting procedure. More realistic uncertainties will be larger owing to a range of systematic effects present in this complex system (see Section \ref{sec:outflows_eta_car} for further discussion).

We find broadly stable results for the higher excitation Balmer lines between H-gamma and H-kappa. For these lines the standard deviations in the resulting parameters are $4.3$ days, $0.02$, $3.7 \kmpers{}$ and $13^\circ$ for $T_{0}$, $e$, $k$ and $\omega$ respectively. We identify some slight degeneracies between two pairs of the parameters: later $T_{0}$ with increases in $\omega$ and higher $e$ with increases in $k$. These slight degeneracies are also present in the original run, as seen by the elliptic covariance in the 2D histograms of these parameter pairs in Figure \ref{fig:ckm_corner}. 

As for H-alpha and H-beta, these lines show larger departures from the original parameter results. We know from Figure \ref{fig:ckm_model_fits} that our CKM model struggles to fit these lines that form in highly extended regions, and so it is not surprising that these lines perform poorly in our cross-validation testing. As previously mentioned, we are not overly concerned by the performance of our CKM model on these two lines due to complicating factors that are present in \etacar{} (see Section \ref{sec:outflows_eta_car} for further discussion).

Further to testing individual lines, we run our CKM model against a subset of Balmer lines, H-gamma to H-kappa, to analyse the sensitivity of the results to the systematic effects most prominent in H-alpha and H-beta. The parameters from this test are shown in the bottom section of Table \ref{tab:ckm_cross_validation}, labelled as H-gamma/kappa. The results show only minimal changes from the original run, demonstrating the resilience of the model to systematic effects when using a larger number of lines.

Overall, we find that it is possible to recover similar orbital solutions regardless of the line used (omitting H-alpha and H-beta). The original run parameters are plotted in Figure \ref{fig:kepler_model_trends}, as red dashed asymptotes, indicating the unified orbital solution our CKM model deduces for \etacar{}, in comparison to the variable Keplerian solutions.

\section{Discussion}
\label{sec:discussion}

We have presented a semi-analytical model, CKM, which takes the first steps towards reconciling the systematic trends observed in different emission lines' radial velocity curves with a consistent orbital solution. In this section we will discuss the nuances of the model in greater detail, as well as the impact the results have on future work involving \etacar{} and other stars with powerful winds.

\subsection{CKM Timescales}
\label{sec:timescales}

The CKM model is built on a simple paradigm intended to encapsulate the dynamics associated with stellar winds out-flowing from binary stars. By formulating the model as a convolution we are able to generate CKM radial velocity curves relatively simply and quickly. This allows us to sample a multidimensional parameter space for the best-fitting solution. However, this description overlooks some physical details; in particular, the star's deviation away from the reference frame in which the wind is launched outwards. This simplification has implications for the radiation field incident on the wind, the gravitational potential felt by the gas and the calculation of wind flow times to line-formation regions.

To check the magnitude of these effects we compare two fundamental timescales associated with the model. First, the orbital timescale, $\tau_{\rm{orbital}}$, defined by
\begin{equation}
\lVert \mathbf{r}(t) + \tau_{\rm{orbital}} \Dot{\mathbf{r}}(t) - \mathbf{r}(t + \tau_{\rm{orbital}}) \rVert = D,
	\label{eq:orbital_timescale}
\end{equation}
where $\mathbf{r}$ is the orbital position, $\Dot{\mathbf{r}}$ is the orbital velocity and $D$ is the deviation distance. $\tau_{\rm{orbital}}$ represents the time it takes for the displacement between the star's future position, based on tangential motion versus orbital motion, to grow to a distance, $D$. Second, the wind flow timescale, $\tau_{\rm{flow}}$, is the time it takes for gas to flow from the sonic point out to a given line-formation radius, $D$. $\tau_{\rm{flow}}$ is calculated by integrating Equation \ref{eq:integrate_velocity_for_time}, resulting in
\begin{equation}
\tau_{\rm{flow}} = \frac{R_{*} (v_{\infty} - v_{0}) \ln (R_{*} v_{0} - v_{\infty} R_{*} + v_{\infty} r) + v_{\infty} r}{v_{\infty}^{2}} \Bigg|_{r=r_{s}}^{r=D}.
	\label{eq:wind_flow_timescale}
\end{equation}

\begin{figure}
	\includegraphics[width=1\columnwidth]{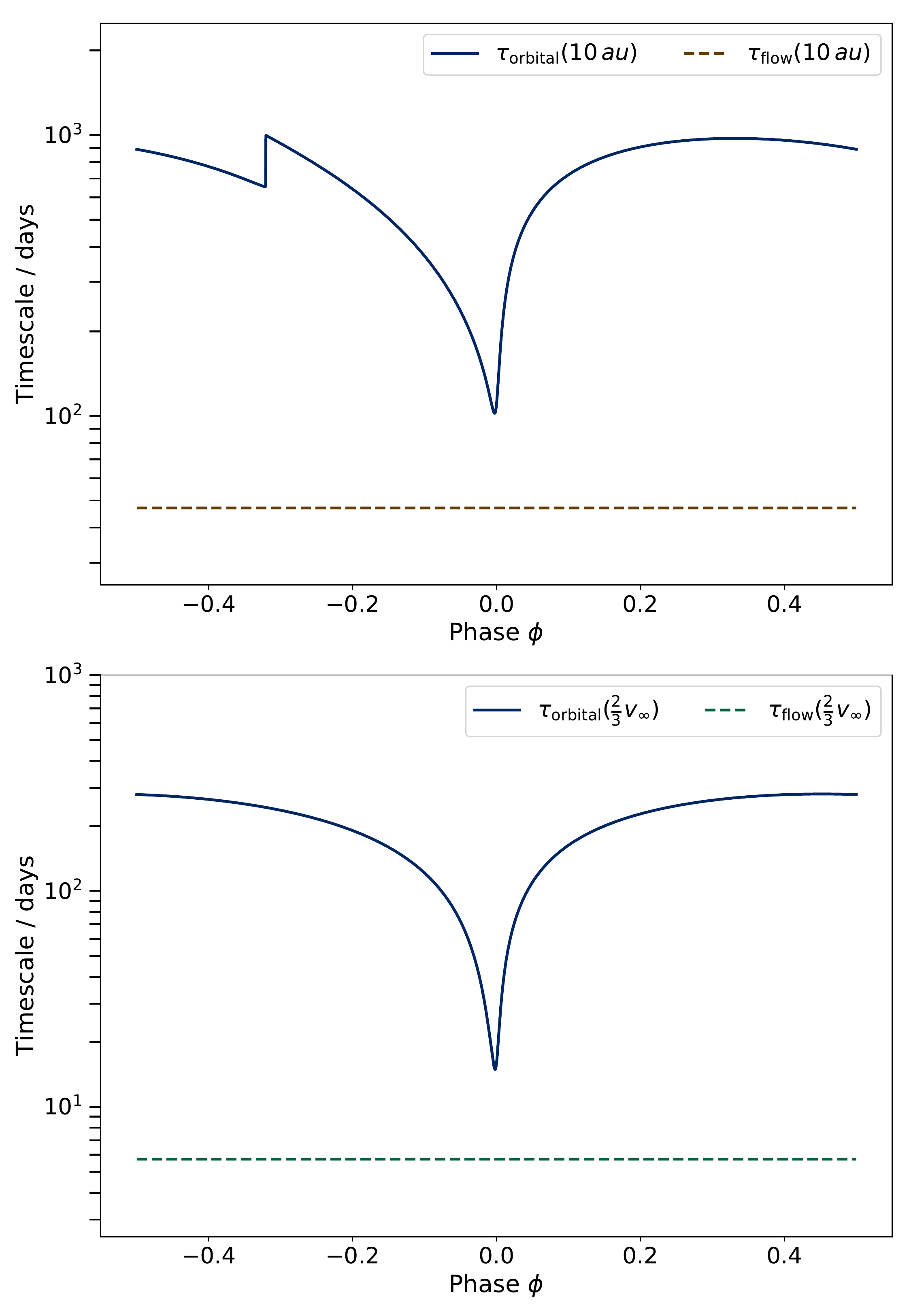}
    \caption{Top panel: timescales comparison of $\tau_{\rm{orbital}}$ (dark blue line) against $\tau_{\rm{flow}}$ (brown dashed line) for a distance scale of $10 \,\rm{AU}$. Bottom panel: timescales comparison of $\tau_{\rm{orbital}}$ (dark blue line) against $\tau_{\rm{accel}}$ (green dashed line) for a distance scale equal to the distance at which the wind has reached $\frac{2}{3} v_{\infty}$. The period of \etacar{} is 2022.7 days.}
    \label{fig:ckm_timescales}
\end{figure}

In Figure \ref{fig:ckm_timescales} we analyse these timescales across an orbital cycle of \etacar{}. In the top panel we compare $\tau_{\rm{orbital}}$ (dark blue line) against $\tau_{\rm{flow}}$ (brown dashed line) for $D=10 \au{}$: a distance comparable to the radius at which an extended line peaks. We find $\tau_{\rm{orbital}} \gg \tau_{\rm{flow}}$ at times around apastron, with the timescales becoming more comparable at periastron due to the highly eccentric orbit. The slight asymmetry in $\tau_{\rm{orbital}}$ is due to the changes in the future orbital path for epochs before or after periastron. The slope to shorter timescales before periastron is shallower; this indicates that the larger accelerations associated with the upcoming periastron cause larger deviations in less time relative to after periastron. We also find a step change in $\tau_{\rm{orbital}}$ at $\phi = -0.32$ ($= 0.68$) which is in an interesting artefact resulting from the sharp changes in motion at periastron. At this phase it takes a sufficiently long time for $D$ to reach $10 \au{}$, such that periastron has passed and the change in direction of the star results in the deviation distance actually decreasing momentarily -- hence the suddenly longer timescales. This analysis shows that changes to the gravitational potential and wind flow times due to CKM not accounting for the star's deviation away from its inertial frame of reference are small.

In the bottom panel of Figure \ref{fig:ckm_timescales} we compare $\tau_{\rm{orbital}}$ (dark blue line) against $\tau_{\rm{flow}}$ (green dashed line) for $D=0.78 \au{}$: the distance it takes the wind to reach $\frac{2}{3} v_{\infty}$. Again, we find $\tau_{\rm{orbital}} \gg \tau_{\rm{accel}}$ at times around apastron, with timescales becoming closer at periastron. The significantly shorter deviation distance, and hence shorter timescales, used in this panel result in $\tau_{\rm{orbital}}$ appearing more symmetric, relative to the top panel, as the future orbital motion has less time to become a factor. This analysis shows that changes to the radiation field incident on the wind are small; and therefore, the line-formation regions and acceleration of the wind will not be greatly affected.

As a result, the timescales considered here establish that our formalism of CKM provides a valid description of the physics associated with powerful winds from binary stars, such as \etacar{}.

\subsection{The outflows of \texorpdfstring{$\eta$}{} car}
\label{sec:outflows_eta_car}

\subsubsection{A latitudinally dependent wind}
\label{sec:latitudinally_dependent_wind}

It may be the case for \etacar{}, and similar stars, that the mass loss is not spherically symmetric. Theoretical work by \citet[][]{Bjorkman1993EquatorialWind} originally showed that above the sonic point the wind should curve towards the equatorial plane and form a disk, so long as it is only subject to radial forces and hence angular momentum is conserved. However, when including both the non-radial line-forces and gravity darkening in models of rapidly rotating luminous stars, the formation of a disk was prevented and the mass flux was actually enhanced in the polar directions \citep[][]{Cranmer1995TheStars, Owocki1996InhibitionWinds}. Furthermore, observations of reflections from \etacar{}'s Homunculus nebula by \citet[][]{Smith2003LatitudedependentCarinae} showed the Balmer lines have faster and deeper P Cygni absorption at higher latitudes, strongly suggestive of a polar enhanced wind.

In our modelling we have used a spherically symmetric wind. It would be possible to include a latitudinal dependence to the wind by calculating the emission-line regions as a function of both radius and polar angle, $\lambda_{L,n}(r, \theta)$, from a 2D stellar atmosphere simulation. The line-formation kernels would then be computed by summing the line contributions over all directions as a function of wind flow time. We have not included this complexity in our model, in part to reduce the number of free parameters, but also because the effects of a faster and denser wind in the polar directions are degenerate (see Appendix \ref{sec:line_formation_kernel_degeneracy}). Hence, any changes to the line-formation kernels will be small.

\subsubsection{The colliding winds}
\label{sec:colliding_winds}

\etacar{} is a colliding wind binary \citep[][]{Pittard1998TheCarinae}. In addition to the powerful primary wind, the companion also emits a strong wind with a mass-loss rate of ${\sim} 10^{-5} \masslossrate{}$ and terminal wind velocity of ${\sim} 2000-3000 \kmpers{}$ \citep[][]{Pittard2002, Parkin2011SpiralingCarinae}. The winds collide where the ram pressures are equal, forming shock fronts in each wind separated by a contact discontinuity. The higher wind momentum flux of the primary wind results in the colliding wind region being shifted towards, and wrapped in a cone-like geometry around, the companion \citep[][]{Okazaki2008ModellingCollision, Parkin2011SpiralingCarinae}.

The cavity caused by the lower density secondary wind may produce systematic effects on the bulk velocities of emission lines. \citet[][]{Madura2012ConstrainingEmission} estimate the semi-major axis of the orbit is ${\sim} 15.4 \au$ (${\sim} 1.5 \au$ binary separation at periastron) and therefore the lower energy Balmer lines, emitted at radii with the greatest overlap with the cavity, will be the most effected.

Given we have calculated $\omega=241^\circ$, the cavity will reduce the velocities away from our line-of-sight (positive velocities) at periastron. This reduction may be delayed to times soon after periastron because of Coriolis effects on the colliding wind geometry \citep[][]{Walder1995SIMULATIONS}. As a result, the observed radial velocity curves may show a reduction in positive velocities in the wake of periastron, appearing as a reduction in $e$, $k$ and $\omega$. This may resolve some of the remaining deviations in our CKM model fitting previously found in Figure \ref{fig:ckm_model_fits} for H-alpha and H-beta.

Outside of the cavity, the colliding winds can produce additional features in the line profiles due to emission in the post-shock gas \citep{Groh2012OnSpectra, Groh2012ACarinae} and the radiation field can influence the ionisation structure of the primary star's wind \citep[][]{Kruip2011ConnectingAlgorithm, Madura2012ConstrainingEmission, Clementel20153DApastron, Clementel2015}. These complications can bias the extraction of radial velocities, as was deduced by \citet[][]{Nielsen2007Interactions} when using the \ion{He}{I} P Cygni absorption as a proxy for orbital motion. In our study we have used emission from a large volume of gas to track the orbital motion; hence, our results may be be less susceptible to some of these systematic effects which only affect specific volumes of the wind on the companions side of the system. Nevertheless, they are source of systematic errors in our results.

\subsubsection{Radiative inhibition and braking vs gravity}
\label{sec:radiative_inhibition_braking_vs_gravity}

When the separation between binary stars is small the radiation field of one star may affect the velocity profile of the other star's wind by radiative inhibition \citep{Stevens1994Stagnation-pointSystems} or braking \citep{Gayley1997SuddenWinds}. Radiative inhibition reduces the initial acceleration of the other star's stellar wind, whilst radiative braking suddenly decelerates the wind as it approaches the companion. Additionally, depending on the ratio of luminosities and masses the countervailing effect of gravity may increase the wind velocities and mass-loss rate \citep{Stevens1994Stagnation-pointSystems}.

\citet[][]{Parkin20093DInhibition} analysed the influence of these effects for \etacar{}. They determined that radiative braking does not occur in either the primary or companion's wind, as the winds collide at their balance point before approaching close enough to the other star. \citet[][]{Parkin20093DInhibition} found that the primary star significantly inhibits the companion's wind, but the primary star's wind is not greatly impacted. Additionally, we compare the companion's gravitational field to the location of the contact discontinuity, along the line between the stars at periastron, and find the winds collide before the companion's gravity becomes a significant factor.

In our work we therefore expect the systematic errors as a result of these effects to be small. Moreover, these calculations were performed for periastron and along the line between the binary star components where the effects are maximal. At phases either side of periastron and elsewhere in the primary wind the influence of these effects will be vastly reduced, further reducing the influence on the emission-line velocities we extracted.

\subsubsection{Time-dependent mass loss}
\label{sec:time_dependent_mass_loss}

There is evidence for \etacar{}'s mass-loss rate increasing immediately following periastron for approximately $80 \days{}$ \citep[]{Corcoran2001TheLoss}. An enhanced mass-loss rate causes emission lines to become more extended as the line-formation regions are dependent on opacity \citep[][]{Hillier1987An50896}. The more extended the emission-line regions, the greater the deviations from the true Keplerian motion the radial velocity curves will show. In this case, these deviations will be exacerbated over the ${\sim} 80 \days{}$ ($0 \lesssim \phi \lesssim 0.04$) following periastron. This may be another reason for the remaining deviations in our CKM model fitting previously found in Figure \ref{fig:ckm_model_fits}.

\subsection{The orbital dynamics of \texorpdfstring{$\eta$}{} Car}
\label{sec:orbital_dynamics_eta_car}

\subsubsection{A comparison of orbital solutions}
\label{sec:comparing_orbital_solutions}

By applying our new CKM model to the Balmer lines in the 2009 periastron GMOS/Gemini data of \etacar{} we have calculated the orbital parameters to be $e=0.91$, $k=69 \kmpers{}$ and $\omega=241^\circ$.

For comparison, previous studies have used a wide range of observational diagnostics, and epochs, to determine the orbital parameters. \citet[][1992 periastron data]{Damineli1997EtaBinary} was the first to propose an orbital solution after taking radial velocity measurements of the Paschen-gamma line, and they calculated the orbital parameters to be $e=0.63$, $k=53 \kmpers{}$ and $\omega=286^\circ$. This solution was updated by \citet[][1992 periastron data]{Davidson1997IsBinary} by including the Paschen-delta and \ion{He}{I} lines and allowing the time of periastron to vary, finding $e=0.8$, $k=65 \kmpers{}$ and $\omega=286^\circ$. \citet[][2003 periastron data]{Nielsen2007Interactions} used \ion{He}{I} P Cygni absorption and calculated $e=0.9$, $k=140 \kmpers{}$ and $\omega=270^\circ$. However, they note the absorption velocities may be influenced by changes in the ionisation regions of \ion{He}{} around periastron, leading to overestimates of $k$. \citet[][2009 \& 2014 periastron data]{Kashi2016ORBITALSYSTEM} made use of \ion{He}{I} and \ion{N}{II} lines, finding $e=0.9$, $k=51 \kmpers{}$ and $\omega=90^\circ$ when assuming the lines form in the companion's wind, in contradiction to the orientation found by most authors.

Our results show good agreement with the prevailing view of \etacar{}'s orbital parameters, having a highly eccentricity binary orientated with the companion on the far side of the primary at periastron with respect to our line-of-sight.

\subsubsection{A comparison of periastron timings}
\label{sec:comparing_periastron_timings}

The results from our CKM model fitting also provide an estimate for the timing of the 2009 periastron (start of cycle number 12 as defined by \citet[][]{Groh2004EtaEvent}), resulting in a value of $T_{0}=2454848$ (JD).

Previously, \citet[][]{Damineli2008TheEvents} derived $T_{0}=2454842.5 \pm 2 \days{}$ (JD) by using the disappearance of the narrow component of \ion{He}{I} 6678 \mbox{\normalfont\AA} line. \citet[][]{Teodoro2016HeMINIMA} derived $T_{0}=2454851.7 \pm 1.3 \days{}$ (JD) by modelling the equivalent width of the \ion{He}{II} 4686 \mbox{\normalfont\AA} line. \citet[][]{Hamaguchi2014X-raySolution} observed the deep minimum in hard X-rays to occur between $2454843$ and $2454859$ (JD).

Our result lies between the values derived from the optical studies and in good agreement with the x-ray minimum. From the eclipse/collapse models of the x-ray light curve \citep[see eg.][]{Pittard1998TheCarinae, Parkin20093DInhibition} we may expect the centre of the deep x-ray minimum, $2454851$ (JD), to occur at superior conjunction. For our derived orbital elements we expect periastron to occur $\sim 3 \days{}$ earlier than superior conjunction, which aligns well with our value of $T_{0}=2454848$ (JD).

\subsubsection{Inferring the masses}
\label{sec:inferring_the_masses}

The new value we have calculated for the semi-amplitude, $k=69 \rm{\,km \, s^{-1}}$, allows us to estimate the stellar masses if we make assumptions about the systems inclination and Eddington fraction. The semi-amplitude can be written in the form
\begin{equation}
k = \frac{29.78 \rm{\,km \, s^{-1}}}{\sqrt[]{1 - e^{2}}} \frac{M_{\rm{B}} \sin i}{\rm{M_\odot}} \bigg( \frac{M_{\rm{A}} + M_{\rm{B}}}{\rm{M_\odot}} \bigg)^{-2/3} \bigg( \frac{P}{1 \rm{yr}} \bigg)^{-1/3},
	\label{eq:semi_amplitude}
\end{equation}
where $M_{\rm{A}}$ and $M_{\rm{B}}$ are the masses of the primary and companion respectively and $i$ is the inclination.

\begin{figure}
	\includegraphics[width=1\columnwidth]{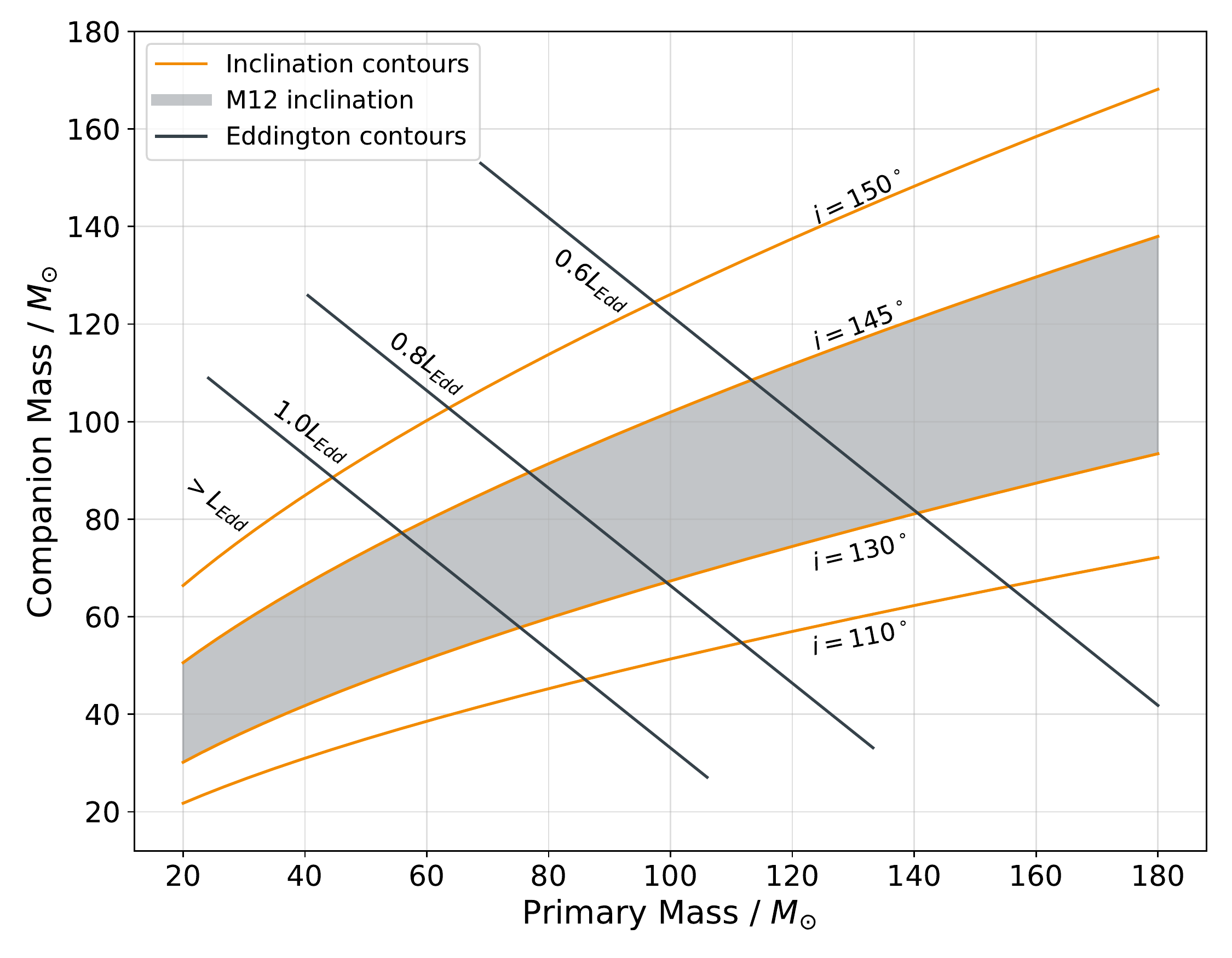}
    \caption{The masses of \etacar{} based on $k=69 \kmpers{}$ for the primary's orbit. The orange lines represent contours of constant inclination. The grey shaded area is the confidence interval for the inclination taken from the \citet[][]{Madura2012ConstrainingEmission} study. The black lines show contours of constant Eddington fraction or total mass.}
    \label{fig:mass_ratio_contours}
\end{figure}

First, only assuming values for $i$ we are able to solve Equation \ref{eq:semi_amplitude} and plot contours of constant inclination on a diagram of mass ratio as shown in Figure \ref{fig:mass_ratio_contours} (orange lines). Modelling by \citet[][]{Madura2012ConstrainingEmission} of the broad [\ion{Fe}{III}] emission had constrained the inclination to the interval $130^\circ < i < 145^\circ$, and we show this confidence interval as the shaded region in Figure \ref{fig:mass_ratio_contours}.

To further constrain the masses requires an estimate for the total mass of the system. Following the work by \citet[][]{Hillier2001CMFGEN}, the Eddington luminosity for a star composed of fully ionised H and He is given by
\begin{equation}
L_{\rm{Edd}} = 3.22 \times 10^{4} \bigg( \frac{N(\rm{H})/N(\rm{He}) + 4)}{N(\rm{H})/N(\rm{He}) + 2} \bigg) \bigg(\frac{M}{\rm{M_{\odot}}}\bigg) \rm{L_{\odot}},
	\label{eq:eddington_mass}
\end{equation}
where $M$ is the Eddington mass of the system and $N(\rm{H})/N(\rm{He})$ is abundance ratio of hydrogen to helium. Using Equation \ref{eq:eddington_mass} we estimate $M$ for different fractions of the Eddington limit, assuming $N(\rm{H})/N(\rm{He}) = 10$ and $L = 5 \times 10^{6} \solarl{}$ \citep[][]{Davidson1997ETAENVIRONMENT}. The Eddington mass may be attributed to either the primary, $M = M_{\rm{A}}$, or the total mass of the system, $M = M_{\rm{A}} + M_{\rm{B}}$. For the latter case the results are plotted as contours of constant Eddington fraction or total mass in Figure \ref{fig:mass_ratio_contours} (black lines).

Examining Figure \ref{fig:mass_ratio_contours} for $i = 137.5^\circ$ and $L=L_{\rm{Edd}}$ we obtain estimates for the stellar masses. For the case where $M = M_{\rm{A}}$ we estimate $M_{\rm{A}} = 133 \solarm{}$ and $M_{\rm{B}} = 93 \solarm{}$, and for the case where $M = M_{\rm{A}} + M_{\rm{B}}$ we estimate $M_{\rm{A}} = 67 \solarm{}$ and $M_{\rm{B}} = 66 \solarm{}$.

\subsection{Application to RMC 140}
\label{sec:application_to_r140}

\begin{figure}
	\includegraphics[width=1\columnwidth]{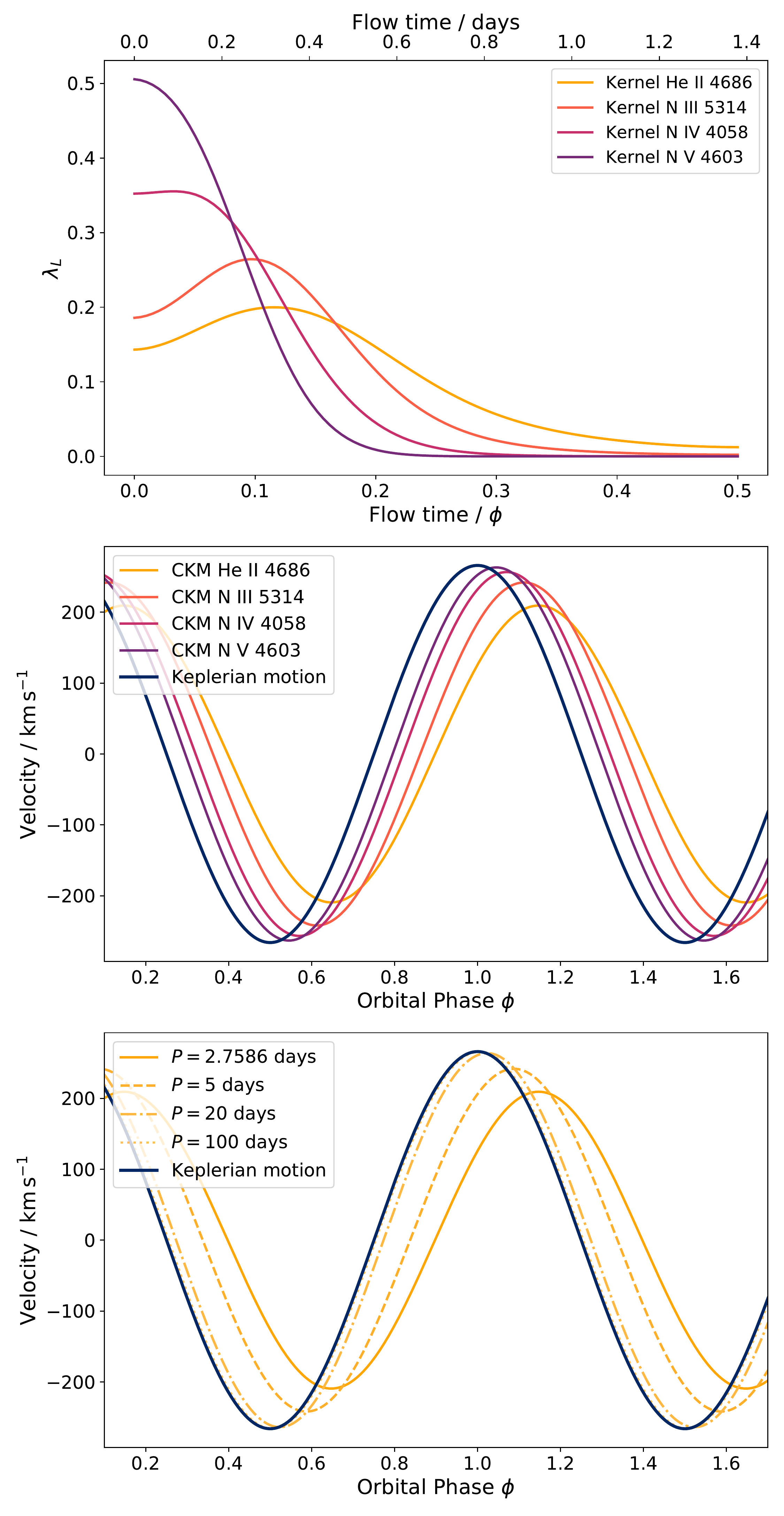}
    \caption{Top panel: line-formation kernels for canonical radial velocity extraction lines for RMC 140b. Each kernel is the linear luminosity density (otherwise known as luminosity per unit radius) as a function of wind flow time beyond the sonic point. Flow time is shown in days (top x-axis) and in units of RMC 140's orbital phase (bottom x axis). Middle panel: CKM deviations from the Keplerian orbit for each line. Bottom panel: deviations from the Keplerian orbit for variations in period for \ion{He}{II} 4686 \mbox{\normalfont\AA}.}
    \label{fig:ckm_canonical_rvlines}
\end{figure}

We investigate the binary system RMC 140 as a secondary case study which is more representative of a larger population of emission-line stars, relative to the esoteric \etacar{}. In particular, we apply our CKM model to the WR star RMC 140b to understand, in principle, the extent to which extended emission may affect the computed orbital solutions of stars with powerful winds.

RMC 140b has literature estimates for the mass-loss rate $\Dot{M} = 1.26 \times 10^{-5} \masslossrate{}$, terminal wind velocity of $v_{\infty} = 1300 \kmpers{}$ and orbital parameters $P=2.7586 \days{}$, $e=0$, $k_{a}=137 \kmpers{}$ and $k_{b}=266 \kmpers{}$ \citep[][]{Shenar2019TheEvolution}. We calculate line-formation kernels, in the same way as described in Section \ref{sec:radiative_transfer}, for a few canonical lines used in extracting radial velocities: \ion{He}{II} 4686 \mbox{\normalfont\AA}, \ion{N}{III} 5314 \mbox{\normalfont\AA}, \ion{N}{IV} 4058 \mbox{\normalfont\AA}, and \ion{N}{V} 4603 \mbox{\normalfont\AA}. The resulting line-formation kernels can be seen in the top panel of Figure \ref{fig:ckm_canonical_rvlines} as a function of wind flow time in days (top axis) and the orbital phase of RMC 140 (bottom axis).

We convolve each line-formation kernel with the literature Keplerian solution to generate a set of CKM curves, plotted in the middle panel of Figure \ref{fig:ckm_canonical_rvlines}. The results show substantial deviations between the Keplerian motion and CKM, with the lines formed furthest from the star showing the largest changes. \ion{N}{V} shows a ${\sim} 5\%$ change in the time of periastron and a ${\sim} 3 \kmpers{}$ decrease in semi-amplitude. The lower energy line, \ion{He}{II}, shows a ${\sim} 15\%$ change in the time of periastron and a ${\sim} 57 \kmpers{}$ decrease in semi-amplitude.

The deviations between our predicted radial velocities for each line and the true Keplerian motion of RMC 140b have important ramifications for future higher accuracy work. Despite using the lines that are commonly adopted for extracting radial velocities we find variations between the resulting motion. This variability highlights the need to account for non-Keplerian effects and carefully consider which emission lines are used when extracting radial velocities to a high degree of accuracy. In the case of RMC 140b these differences are being driven by the size of the line-formation kernels relative to the orbital period. The short period of $P=2.7586 \days{}$ exacerbates the time-averaging effects of CKM. To test the sensitivity of the deviations on the period we create a set of CKM curves for a range of artificial period lengths between the original $2.7586 \days{}$ and $100 \days{}$. The results are shown for the \ion{He}{II} line in the bottom panel of Figure \ref{fig:ckm_canonical_rvlines}. The CKM curve for the original $2.7586$ day period is identical between the middle and bottom panels. However, as the period length increases, the deviations between the Keplerian motion and CKM curves incrementally decreases. At a period of ${\sim} 100 \days{}$ the differences become negligible.

In general, the expected size of the deviations between radial velocity curves and the true Keplerian motion depends on the interplay between the line-formation kernels and the orbit. The more extended the line emission region relative to the orbital timescales, the larger the deviations. The emission region is more extended for higher mass-loss rate stars, and the wind flow times are longer for stars with slower winds. As a consequence, LBVs showcase the deviations to maximal effect and other types of massive evolved stars are good candidates for observing further cases of this effect.

\section{Summary and conclusions}
\label{sec:summary_and_conclusions}
We have investigated discrepancies in radial velocity data between spectral lines of varying excitation emitted from binary stars with powerful winds. We have presented a new semi-analytical model, convolutional Keplerian motion (CKM), which encapsulates both the star's orbital motion and the propagation of the wind, accounting for the time delay to the location where the photons are emitted. Our work is summarised as follows:
\begin{enumerate}
  \item We made use of archival Gemini/GMOS observations of \etacar{} from across the 2009 periastron. We employed a multi-Gaussian fitting algorithm to decompose the complex line profiles of the Balmer series into their constituent dynamical components and extracted radial velocities from the primary star's wind emission.
  \item As an instructive exercise we fit the radial velocity observations with Keplerian models, finding monotonic trends in the orbital parameters $T_0$, $e$ and $k$ when ordering the lines by their transition energy: different lines produce different orbital solutions.
  \item To reconcile the discrepancies in the orbital parameters between lines we formulated a semi-analytical model, CKM, which encapsulates the dynamics associated with lines formed in extended regions, encoding the integrated velocity field of the out-flowing gas in terms of a convolution of past motion due to the finite flow speed of the wind.
  \item We fit our CKM model to 10 of \etacar{}'s Balmer lines, from H-alpha to H-kappa, concurrently. The best fitting orbital parameters are $T_{0}=2454848$ (JD), $e=0.91$, $k=69 \kmpers{}$ and $\omega=241^\circ$. Cross-validation showed stability in these parameters for the higher excitation Balmer lines, but we found systematic effects, most likely due to the influence of the companion, are apparent in H-alpha and H-beta. Overall, this model is a first step towards being able to generate a consistent set of Keplerian parameters independent of the emission line used.
  \item We applied our CKM model to a more typical case: the WR star in the binary RMC 140. Despite selecting the canonical lines used for extracting radial velocities (\ion{He}{II} and \ion{N}{III/IV/V}), deviations still exist between the published Keplerian orbital parameters and the predicted radial velocities: an important correction for future higher accuracy work. We found these deviations are driven by the short orbital period in RMC 140.  
  \item More generally, the magnitude of the deviations between radial velocity curves and the true Keplerian motion depends on the interplay between the line-formation kernels and the orbit. The more extended the line emission regions relative to the orbital timescales, the larger the deviations.
\end{enumerate}
The model presented in this study shows good support from the observations of \etacar{}'s emission-line dynamics. Future work could involve radiative-hydrodynamical  simulations, ideally in 3D, to fully investigate the complexities, timescales and systematics present. Further observational datasets of other binary stars exhibiting powerful winds are also required to help strengthen this research in the future. Nevertheless, the work presented here is an important proof of concept and demonstrates the need to implement models, such as CKM, in order to uncover the true orbital motion of binary stars with powerful winds.

\section*{Acknowledgements}
\label{sec:acknowledgements}
This research has made use of the data archive for Eta Carinae which is available online at \url{http://etacar.umn.edu}. The archive is supported by the University of Minnesota and the Space Telescope Science Institute under contract with NASA. We would like to thanks the anonymous referee for a helpful report that improved the quality of the paper. We would also like to thank R. Hirai, S. Owocki and P. Podsiadlowski for helpful discussions. We gratefully acknowledge the use of the following software: \texttt{CMFGEN} \citep[][]{Hillier1998TheOutflows, Hillier2001CMFGEN}, \texttt{SciPy} \citep[][]{Jones2001SciPy:Python}, NumPy \citep[][]{VanDerWalt2011TheComputation}, Pandas \citep[][]{McKinney2010DataPython}, emcee \citep[][]{Foreman-Mackey2013Emcee:Hammer} and Matplotlib \citep[][]{Hunter2007Matplotlib:Environment}. Software developed for this research is open-source and freely available at \url{https://github.com/DavoGrant/}.




\bibliographystyle{mnras}
\bibliography{references} 

\begin{thebibliography}{}
\makeatletter
\relax
\def\mn@urlcharsother{\let\do\@makeother \do\$\do\&\do\#\do\^\do\_\do\%\do\~}
\def\mn@doi{\begingroup\mn@urlcharsother \@ifnextchar [ {\mn@doi@}
  {\mn@doi@[]}}
\def\mn@doi@[#1]#2{\def\@tempa{#1}\ifx\@tempa\@empty \href
  {http://dx.doi.org/#2} {doi:#2}\else \href {http://dx.doi.org/#2} {#1}\fi
  \endgroup}
\def\mn@eprint#1#2{\mn@eprint@#1:#2::\@nil}
\def\mn@eprint@arXiv#1{\href {http://arxiv.org/abs/#1} {{\tt arXiv:#1}}}
\def\mn@eprint@dblp#1{\href {http://dblp.uni-trier.de/rec/bibtex/#1.xml}
  {dblp:#1}}
\def\mn@eprint@#1:#2:#3:#4\@nil{\def\@tempa {#1}\def\@tempb {#2}\def\@tempc
  {#3}\ifx \@tempc \@empty \let \@tempc \@tempb \let \@tempb \@tempa \fi \ifx
  \@tempb \@empty \def\@tempb {arXiv}\fi \@ifundefined
  {mn@eprint@\@tempb}{\@tempb:\@tempc}{\expandafter \expandafter \csname
  mn@eprint@\@tempb\endcsname \expandafter{\@tempc}}}

\bibitem[\protect\citeauthoryear{Aitken}{Aitken}{1964}]{Aitken1964TheStars}
Aitken R.~G.,  1964, {The Binary Stars}.
Dover Publication, New York

\bibitem[\protect\citeauthoryear{Allen \& Hillier}{Allen \&
  Hillier}{1993}]{Allen1993TheCarinae}
Allen D.,  Hillier D.,  1993, \mn@doi [Publications of the Astronomical Society
  of Australia] {10.1017/s1323358000025996}, 10, 338

\bibitem[\protect\citeauthoryear{Beals}{Beals}{1929}]{Beals1929OnEmission}
Beals C.~S.,  1929, MNRAS, 90, 202

\bibitem[\protect\citeauthoryear{Belczynski, Kalogera  \& Bulik}{Belczynski
  et~al.}{2002}]{Belczynski2002AProperties}
Belczynski K.,  Kalogera V.,   Bulik T.,  2002, \mn@doi [The Astrophysical
  Journal] {10.1086/340304}, 572, 407

\bibitem[\protect\citeauthoryear{Bjorkman \& Cassinelli}{Bjorkman \&
  Cassinelli}{1993}]{Bjorkman1993EquatorialWind}
Bjorkman J.~E.,  Cassinelli J.~P.,  1993, \mn@doi [The Astrophysical Journal]
  {10.1086/172676}, 409, 429

\bibitem[\protect\citeauthoryear{Blundell, Bowler  \& Schmidtobreick}{Blundell
  et~al.}{2007}]{Blundell2007FluctuationsTimescales}
Blundell K.~M.,  Bowler M.~G.,   Schmidtobreick L.,  2007, \mn@doi [Astronomy
  {\&} Astrophysics] {10.1051/0004-6361:20077924}, 474, 903

\bibitem[\protect\citeauthoryear{Blundell, Bowler  \& Schmidtobreick}{Blundell
  et~al.}{2008}]{Blundell2008SSMass}
Blundell K.~M.,  Bowler M.~G.,   Schmidtobreick L.,  2008, \mn@doi [The
  Astrophysical Journal] {10.1086/588027}, 678, L47

\bibitem[\protect\citeauthoryear{Blundell, Schmidtobreick  \&
  Trushkin}{Blundell et~al.}{2011}]{Blundell2011SS433sFlare}
Blundell K.~M.,  Schmidtobreick L.,   Trushkin S.,  2011, \mn@doi [Monthly
  Notices of the Royal Astronomical Society]
  {10.1111/j.1365-2966.2011.18785.x}, 417, 2401

\bibitem[\protect\citeauthoryear{Bowen}{Bowen}{1928}]{Bowen1928THENEBULAE}
Bowen I.~S.,  1928, The Astrophysical Journal, 67

\bibitem[\protect\citeauthoryear{Branch, Coleman  \& Li}{Branch
  et~al.}{1999}]{Branch1999AProblems}
Branch M.~A.,  Coleman T.~F.,   Li Y.,  1999, \mn@doi [SIAM Journal on
  Scientific Computing] {10.1137/s1064827595289108}, 21, 1

\bibitem[\protect\citeauthoryear{Castor, Abbot  \& Klein}{Castor
  et~al.}{1975}]{Castor1975Radiation-drivenStars}
Castor J.,  Abbot D.,   Klein R.,  1975, The Astrophysical Journal, 195, 157

\bibitem[\protect\citeauthoryear{Clementel, Madura, Kruip, Paardekooper  \&
  Gull}{Clementel et~al.}{2015a}]{Clementel20153DApastron}
Clementel N.,  Madura T.~I.,  Kruip C. J.~H.,  Paardekooper J.-P.,   Gull
  T.~R.,  2015a, \mn@doi [Monthly Notices of the Royal Astronomical Society]
  {10.1093/mnras/stu2614}, 447, 2445

\bibitem[\protect\citeauthoryear{Clementel, Madura, Kruip  \&
  Paardekooper}{Clementel et~al.}{2015b}]{Clementel2015}
Clementel N.,  Madura T.~I.,  Kruip C.~J.,   Paardekooper J.~P.,  2015b,
  \mn@doi [Monthly Notices of the Royal Astronomical Society]
  {10.1093/mnras/stv696}, 450, 1388

\bibitem[\protect\citeauthoryear{Corcoran, Ishibashi, Swank  \& Petre}{Corcoran
  et~al.}{2001}]{Corcoran2001TheLoss}
Corcoran M.~F.,  Ishibashi K.,  Swank J.~H.,   Petre R.,  2001, \mn@doi [The
  Astrophysical Journal] {10.1086/318416}, 547, 1034

\bibitem[\protect\citeauthoryear{Cranmer \& Owocki}{Cranmer \&
  Owocki}{1995}]{Cranmer1995TheStars}
Cranmer S.~R.,  Owocki S.~P.,  1995, \mn@doi [The Astrophysical Journal]
  {10.1086/175272}, 440, 308

\bibitem[\protect\citeauthoryear{Currie et~al.,}{Currie
  et~al.}{1996}]{Currie1996AstrometricTelescope}
Currie D.~G.,  et~al., 1996, \mn@doi [The Astronomical Journal]
  {10.1086/118083}, 112, 1115

\bibitem[\protect\citeauthoryear{Damineli, Conti, Lopesb  \& Van
  Den~Heuvel}{Damineli et~al.}{1997}]{Damineli1997EtaBinary}
Damineli A.,  Conti P.~S.,  Lopesb D.~F.,   Van Den~Heuvel E. P.~J.,  1997, NEW
  ASTRONOMY EISEYIER New Astronomy, 2, 107

\bibitem[\protect\citeauthoryear{Damineli et~al.,}{Damineli
  et~al.}{2008}]{Damineli2008TheEvents}
Damineli A.,  et~al., 2008, \mn@doi [Monthly Notices of the Royal Astronomical
  Society] {10.1111/j.1365-2966.2007.12815.x}, 384, 1649

\bibitem[\protect\citeauthoryear{Davidson}{Davidson}{1997}]{Davidson1997IsBinary}
Davidson K.,  1997, \mn@doi [New Astronomy] {10.1016/S1384-1076(97)00028-6}, 2,
  387

\bibitem[\protect\citeauthoryear{Davidson \& Humphreys}{Davidson \&
  Humphreys}{1997}]{Davidson1997ETAENVIRONMENT}
Davidson K.,  Humphreys R.~M.,  1997, \mn@doi [Annual Review of Astronomy and
  Astrophysics] {10.1146/annurev.astro.35.1.1}, 35, 1

\bibitem[\protect\citeauthoryear{Davidson, Ishibashi, Gull, Humphreys  \&
  Smith}{Davidson et~al.}{2000}]{Davidson2000}
Davidson K.,  Ishibashi K.,  Gull T.~R.,  Humphreys R.~M.,   Smith N.,  2000,
  \mn@doi [The Astrophysical Journal] {10.1086/312502}, 530, 107

\bibitem[\protect\citeauthoryear{Davidson, Mehner, Humphreys, Martin  \&
  Ishibashi}{Davidson et~al.}{2015}]{Davidson2015ETAFeatures}
Davidson K.,  Mehner A.,  Humphreys R.~M.,  Martin J.~C.,   Ishibashi K.,
  2015, \mn@doi [Astrophysical Journal Letters] {10.1088/2041-8205/801/1/L15},
  801, L15

\bibitem[\protect\citeauthoryear{Dorland, Currie  \& Hajian}{Dorland
  et~al.}{2004}]{Dorland2004Did1941}
Dorland B.~N.,  Currie D.~G.,   Hajian A.~R.,  2004, \mn@doi [The Astronomical
  Journal] {10.1086/380941}, 127, 1052

\bibitem[\protect\citeauthoryear{Ebbets}{Ebbets}{1982}]{Ebbets1982TheSupergiants}
Ebbets D.,  1982, \mn@doi [The Astrophysical Journal Supplement Series]
  {10.1086/190783}, 48, 399

\bibitem[\protect\citeauthoryear{Evans \& Taylor}{Evans \&
  Taylor}{2011}]{Evans2011TheOverview}
Evans C.~J.,  Taylor W.~D.,  2011, \mn@doi [Astronomy and Astrophysics]
  {10.1051/0004-6361/201116782}, 530

\bibitem[\protect\citeauthoryear{Foreman-Mackey, Hogg, Lang  \&
  Goodman}{Foreman-Mackey et~al.}{2013}]{Foreman-Mackey2013Emcee:Hammer}
Foreman-Mackey D.,  Hogg D.~W.,  Lang D.,   Goodman J.,  2013, Technical
  report, {emcee: The MCMC Hammer}, \url
  {http://dan.iel.fm/emceeundertheMITLicense.}.
Center for Cosmology and Particle Physics, Department of Physics, New York
  University, \url {http://dan.iel.fm/emceeundertheMITLicense.}

\bibitem[\protect\citeauthoryear{Gayley, Owocki  \& Cranmer}{Gayley
  et~al.}{1997}]{Gayley1997SuddenWinds}
Gayley K.~G.,  Owocki S.~P.,   Cranmer S.~R.,  1997, \mn@doi [The Astrophysical
  Journal] {10.1086/303573}, pp 786--797

\bibitem[\protect\citeauthoryear{Griffiths}{Griffiths}{2007}]{Griffiths2007IntroductionElectrodynamics}
Griffiths D.~J.,  2007, {Introduction to Electrodynamics}.
Pearson Education, Dorling Kindersley

\bibitem[\protect\citeauthoryear{Groh \& Damineli}{Groh \&
  Damineli}{2004}]{Groh2004EtaEvent}
Groh J.,  Damineli A.,  2004, Information Bulletin on Variable Stars, 5492

\bibitem[\protect\citeauthoryear{Groh, Hillier, Madura  \& Weigelt}{Groh
  et~al.}{2012a}]{Groh2012OnSpectra}
Groh J.~H.,  Hillier D.~J.,  Madura T.~I.,   Weigelt G.,  2012a, \mn@doi
  [Monthly Notices of the Royal Astronomical Society]
  {10.1111/j.1365-2966.2012.20984.x}, 423, 1623

\bibitem[\protect\citeauthoryear{Groh, Madura, Hillier, Kruip  \& Weigelt}{Groh
  et~al.}{2012b}]{Groh2012ACarinae}
Groh J.~H.,  Madura T.~I.,  Hillier D.~J.,  Kruip C.~J.,   Weigelt G.,  2012b,
  \mn@doi [Astrophysical Journal Letters] {10.1088/2041-8205/759/1/L2}, 759, 2

\bibitem[\protect\citeauthoryear{Hamaguchi et~al.,}{Hamaguchi
  et~al.}{2014}]{Hamaguchi2014X-raySolution}
Hamaguchi K.,  et~al., 2014, \mn@doi [Astrophysical Journal]
  {10.1088/0004-637X/784/2/125}, 784, 125

\bibitem[\protect\citeauthoryear{Hillier}{Hillier}{1987}]{Hillier1987An50896}
Hillier D.~J.,  1987, The Astrophysical Journal Supplement Series, 9, 965

\bibitem[\protect\citeauthoryear{Hillier \& Lanz}{Hillier \&
  Lanz}{2001}]{Hillier2001CMFGEN}
Hillier D.~J.,  Lanz T.,  2001, ASP Conference Series, 247

\bibitem[\protect\citeauthoryear{Hillier \& Miller}{Hillier \&
  Miller}{1998}]{Hillier1998TheOutflows}
Hillier D.~J.,  Miller D.~L.,  1998, \mn@doi [Astrophysical Journal]
  {10.1086/305350}, 496, 407

\bibitem[\protect\citeauthoryear{Hillier, Davidson, Ishibashi  \& Gull}{Hillier
  et~al.}{2001}]{Hillier2001}
Hillier D.~J.,  Davidson K.,  Ishibashi K.,   Gull T.,  2001, \mn@doi [The
  Astrophysical Journal] {10.1086/320948}, 553, 837

\bibitem[\protect\citeauthoryear{Hillier et~al.,}{Hillier
  et~al.}{2006}]{Hillier2006TheCarinae}
Hillier D.~J.,  et~al., 2006, \mn@doi [The Astrophysical Journal]
  {10.1086/501225}, 642, 1098

\bibitem[\protect\citeauthoryear{Humphreys, Smith  \& Davidson}{Humphreys
  et~al.}{1999}]{Humphreys1999EtaVariables}
Humphreys R.~M.,  Smith N.,   Davidson K.,  1999, Publications of the
  Astronomical Society of the Pacific, 111, 1124

\bibitem[\protect\citeauthoryear{Hunter}{Hunter}{2007}]{Hunter2007Matplotlib:Environment}
Hunter J.~D.,  2007, \mn@doi [Computing in Science and Engineering]
  {10.1109/MCSE.2007.55}, 9, 99

\bibitem[\protect\citeauthoryear{Jones, Oliphant  \& Peterson}{Jones
  et~al.}{2001}]{Jones2001SciPy:Python}
Jones E.,  Oliphant T.,   Peterson P.,  2001, {SciPy: Open source scientific
  tools for Python}

\bibitem[\protect\citeauthoryear{Kaper, Henrichs  \& Nichols}{Kaper
  et~al.}{1996}]{Kaper1996Long-Stars}
Kaper L.,  Henrichs H.~F.,   Nichols J.~S.,  1996, \mn@doi [Astronomy and
  Astrophysics Supplement Series] {10.1051/aas:1996113}, 116, 257

\bibitem[\protect\citeauthoryear{Kashi \& Soker}{Kashi \&
  Soker}{2016}]{Kashi2016ORBITALSYSTEM}
Kashi A.,  Soker N.,  2016, \mn@doi [The Astrophysical Journal]
  {10.3847/0004-637X/825/2/105}, 825, 105

\bibitem[\protect\citeauthoryear{Kolmogorov}{Kolmogorov}{1933}]{Kolmogorov1933NoTitle}
Kolmogorov A.~N.,  1933, Giornale dell'Instituto Italiano degli Attuari, 4, 83

\bibitem[\protect\citeauthoryear{Konacki}{Konacki}{2005}]{Konacki2005PrecisionCell}
Konacki M.,  2005, \mn@doi [The Astrophysical Journal] {10.1086/429880}, 626,
  431

\bibitem[\protect\citeauthoryear{Konacki, Muterspaugh, Kulkarni  \&
  He{\l}lminiak}{Konacki et~al.}{2010}]{Konacki2010High-precisionHD210027}
Konacki M.,  Muterspaugh M.~W.,  Kulkarni S.~R.,   He{\l}lminiak K.~G.,  2010,
  \mn@doi [Astrophysical Journal] {10.1088/0004-637X/719/2/1293}, 719, 1293

\bibitem[\protect\citeauthoryear{Kruip}{Kruip}{2011}]{Kruip2011ConnectingAlgorithm}
Kruip C.,  2011, PhD thesis, Univ. Lieden

\bibitem[\protect\citeauthoryear{Kuhi}{Kuhi}{1973}]{Kuhi1973Wolf-RayetStratification}
Kuhi L.~V.,  1973, The Astrophysical Journal, 180, 783

\bibitem[\protect\citeauthoryear{MacGregor, Hartmann  \& Raymond}{MacGregor
  et~al.}{1979}]{MacGregor1979RadiativeStars}
MacGregor K.~B.,  Hartmann L.,   Raymond J.~C.,  1979, \mn@doi [The
  Astrophysical Journal] {10.1086/157213}, 231, 514

\bibitem[\protect\citeauthoryear{Madura, Gull, Owocki, Groh, Okazaki  \&
  Russell}{Madura et~al.}{2012}]{Madura2012ConstrainingEmission}
Madura T.~I.,  Gull T.~R.,  Owocki S.~P.,  Groh J.~H.,  Okazaki A.~T.,
  Russell C.~M.,  2012, \mn@doi [Monthly Notices of the Royal Astronomical
  Society] {10.1111/j.1365-2966.2011.20165.x}, 420, 2064

\bibitem[\protect\citeauthoryear{Marchenko et~al.,}{Marchenko
  et~al.}{2003}]{Marchenko2003The140}
Marchenko S.~V.,  et~al., 2003, \mn@doi [The Astrophysical Journal]
  {10.1086/378154}, 596, 1295

\bibitem[\protect\citeauthoryear{McKinney}{McKinney}{2010}]{McKinney2010DataPython}
McKinney W.,  2010, in Proceedings of the 9th Python in Science Conference. pp
  51--56, \url
  {http://conference.scipy.org/proceedings/scipy2010/mckinney.html}

\bibitem[\protect\citeauthoryear{Mehner, Davidson, Martin, Humphreys, Ishibashi
   \& Ferland}{Mehner et~al.}{2011}]{Mehner2011CriticalEvent}
Mehner A.,  Davidson K.,  Martin J.~C.,  Humphreys R.~M.,  Ishibashi K.,
  Ferland G.~J.,  2011, \mn@doi [Astrophysical Journal]
  {10.1088/0004-637X/740/2/80}, 740, 80

\bibitem[\protect\citeauthoryear{Moffat, Niemela, Phillips, Chu  \&
  Seggewiss}{Moffat et~al.}{1987}]{Moffat1987TheSurroundings}
Moffat A. F.~J.,  Niemela V.~S.,  Phillips M.~M.,  Chu Y.-H.,   Seggewiss W.,
  1987, \mn@doi [The Astrophysical Journal] {10.1086/164906}, 312, 612

\bibitem[\protect\citeauthoryear{Nielsen, Corcoran, Gull, Hillier, Hamaguchi,
  Ivarsson  \& Lindler}{Nielsen et~al.}{2007}]{Nielsen2007Interactions}
Nielsen K.~E.,  Corcoran M.~F.,  Gull T.~R.,  Hillier D.~J.,  Hamaguchi K.,
  Ivarsson S.,   Lindler D.~J.,  2007, \mn@doi [The Astrophysical Journal]
  {10.1086/513006}, 660, 669

\bibitem[\protect\citeauthoryear{Okazaki, Owocki, Russell  \& Corcoran}{Okazaki
  et~al.}{2008}]{Okazaki2008ModellingCollision}
Okazaki A.~T.,  Owocki S.~P.,  Russell C.~M.,   Corcoran M.~F.,  2008, \mn@doi
  [Monthly Notices of the Royal Astronomical Society: Letters]
  {10.1111/j.1745-3933.2008.00496.x}, 388, 39

\bibitem[\protect\citeauthoryear{Oskinova, Feldmeier  \& Hamann}{Oskinova
  et~al.}{2006}]{Oskinova2006High-resolutionStars}
Oskinova L.~M.,  Feldmeier A.,   Hamann W.~R.,  2006, \mn@doi [Monthly Notices
  of the Royal Astronomical Society] {10.1111/j.1365-2966.2006.10858.x}, 372,
  313

\bibitem[\protect\citeauthoryear{Owocki, Castor  \& Rybicki}{Owocki
  et~al.}{1988}]{Owocki1988Time-dependentModel}
Owocki S.~P.,  Castor J.~I.,   Rybicki G.~B.,  1988, \mn@doi [The Astrophysical
  Journal] {10.1086/166977}, 335, 914

\bibitem[\protect\citeauthoryear{Owocki, Cranmer  \& Gayley}{Owocki
  et~al.}{1996}]{Owocki1996InhibitionWinds}
Owocki S.~P.,  Cranmer S.~R.,   Gayley K.~G.,  1996, \mn@doi [The Astrophysical
  Journal] {10.1086/310372}, 472, L115

\bibitem[\protect\citeauthoryear{Owocki, Gayley  \& Shaviv}{Owocki
  et~al.}{2004}]{Owocki2004ALimit}
Owocki S.~P.,  Gayley K.~G.,   Shaviv N.~J.,  2004, \mn@doi [The Astrophysical
  Journal] {10.1086/424910}, 616, 525

\bibitem[\protect\citeauthoryear{Parkin, Pittard, Corcoran, Hamaguchi  \&
  Stevens}{Parkin et~al.}{2009}]{Parkin20093DInhibition}
Parkin E.~R.,  Pittard J.~M.,  Corcoran M.~F.,  Hamaguchi K.,   Stevens I.~R.,
  2009, \mn@doi [Monthly Notices of the Royal Astronomical Society]
  {10.1111/j.1365-2966.2009.14475.x}, 394, 1758

\bibitem[\protect\citeauthoryear{Parkin, Pittard, Corcoran  \&
  Hamaguchi}{Parkin et~al.}{2011}]{Parkin2011SpiralingCarinae}
Parkin E.~R.,  Pittard J.~M.,  Corcoran M.~F.,   Hamaguchi K.,  2011, \mn@doi
  [Astrophysical Journal] {10.1088/0004-637X/726/2/105}, 726, 105

\bibitem[\protect\citeauthoryear{Pietrzy{\'{n}}ski et~al.,}{Pietrzy{\'{n}}ski
  et~al.}{2013}]{Pietrzynski2013AnCent}
Pietrzy{\'{n}}ski G.,  et~al., 2013, \mn@doi [Nature] {10.1038/nature11878},
  495, 76

\bibitem[\protect\citeauthoryear{Pittard \& Corcoran}{Pittard \&
  Corcoran}{2002}]{Pittard2002}
Pittard J.~M.,  Corcoran M.~F.,  2002, \mn@doi [A{\&}A]
  {10.1051/0004-6361:20020025}, 383, 636

\bibitem[\protect\citeauthoryear{Pittard, Stevens, Corcoran  \&
  Ishibashi}{Pittard et~al.}{1998}]{Pittard1998TheCarinae}
Pittard J.~M.,  Stevens I.~R.,  Corcoran M.~F.,   Ishibashi K.,  1998, \mn@doi
  [Monthly Notices of the Royal Astronomical Society]
  {10.1046/j.1365-8711.1998.01962.x}, 299, L5

\bibitem[\protect\citeauthoryear{Richardson, Gies, Henry, Fernndez-Lajs  \&
  Okazaki}{Richardson et~al.}{2010}]{Richardson2010TheEvent}
Richardson N.~D.,  Gies D.~R.,  Henry T.~J.,  Fernndez-Lajs E.,   Okazaki
  A.~T.,  2010, \mn@doi [Astronomical Journal] {10.1088/0004-6256/139/4/1534},
  139, 1534

\bibitem[\protect\citeauthoryear{Richardson, Gies, Gull, Moffat  \&
  St-Jean}{Richardson et~al.}{2015}]{Richardson2015TheEvent}
Richardson N.~D.,  Gies D.~R.,  Gull T.~R.,  Moffat A.~F.,   St-Jean L.,  2015,
  \mn@doi [Astronomical Journal] {10.1088/0004-6256/150/4/109}, 150

\bibitem[\protect\citeauthoryear{Riener, Kainulainen, Henshaw, Orkisz, Murray
  \& Beuther}{Riener et~al.}{2019}]{Riener2019GAUSSPY+:Spectra}
Riener M.,  Kainulainen J.,  Henshaw J.~D.,  Orkisz J.~H.,  Murray C.~E.,
  Beuther H.,  2019, \mn@doi [Astronomy {\&} Astrophysics]
  {10.1051/0004-6361/201935519}, 628, A78

\bibitem[\protect\citeauthoryear{Ruiz, Melnick  \& Ortiz}{Ruiz
  et~al.}{1984}]{Ruiz1984TimeCarinae}
Ruiz M.~T.,  Melnick J.,   Ortiz P.,  1984, \mn@doi [The Astrophysical Journal]
  {10.1086/184356}, 285, L19

\bibitem[\protect\citeauthoryear{Saar, Butler  \& Marcy}{Saar
  et~al.}{1998}]{Saar1998MagneticSurvey}
Saar S.~H.,  Butler R.~P.,   Marcy G.~W.,  1998, \mn@doi [The Astrophysical
  Journal] {10.1086/311325}, 498, L153

\bibitem[\protect\citeauthoryear{Sana et~al.,}{Sana
  et~al.}{2012}]{Sana2012BinaryStars}
Sana H.,  et~al., 2012, \mn@doi [Science] {10.1126/science.1223344}, 337, 444

\bibitem[\protect\citeauthoryear{Savitzky \& Golay}{Savitzky \&
  Golay}{1964}]{Savitzky1964SmoothingProcedures}
Savitzky A.,  Golay M.,  1964, \mn@doi [Analytical Chemistry]
  {10.1021/ac60214a047}, 36, 1627

\bibitem[\protect\citeauthoryear{Shenar et~al.,}{Shenar
  et~al.}{2019}]{Shenar2019TheEvolution}
Shenar T.,  et~al., 2019, \mn@doi [Astronomy {\&} Astrophysics]
  {10.1051/0004-6361/201935684}

\bibitem[\protect\citeauthoryear{Shylaja}{Shylaja}{1986}]{Shylaja1986SpectrophotometricCephei}
Shylaja B.~S.,  1986, \mn@doi [Journal of Astrophysics and Astronomy]
  {10.1007/BF02714209}, 7, 171

\bibitem[\protect\citeauthoryear{Shylaja}{Shylaja}{1987}]{Shylaja1987TheBinaries}
Shylaja B.~S.,  1987, \mn@doi [Journal of Astrophysics and Astronomy]
  {10.1007/BF02714316}, 8, 183

\bibitem[\protect\citeauthoryear{Smirnov}{Smirnov}{1939}]{Smirnov1939OnSamples}
Smirnov N.~V.,  1939, Bull. Math. Univ. Moscou, 2, 3

\bibitem[\protect\citeauthoryear{Smith}{Smith}{2006}]{Smith2006TheCarinae}
Smith N.,  2006, \mn@doi [The Astrophysical Journal] {10.1086/503766}, 644,
  1151

\bibitem[\protect\citeauthoryear{Smith}{Smith}{2014}]{Smith2014MassStars}
Smith N.,  2014, \mn@doi [Annual Review of Astronomy and Astrophysics]
  {10.1146/annurev-astro-081913-040025}, 52, 487

\bibitem[\protect\citeauthoryear{Smith, Gehrz, Hinz, Hoffmann, Hora, Mamajek
  \& Meyer}{Smith et~al.}{2003a}]{Smith2003MassCarinae}
Smith N.,  Gehrz R.~D.,  Hinz P.~M.,  Hoffmann W.~F.,  Hora J.~L.,  Mamajek
  E.~E.,   Meyer M.~R.,  2003a, \mn@doi [The Astronomical Journal]
  {10.1016/0965-1748(95)00012-K}, 125, 1458

\bibitem[\protect\citeauthoryear{Smith, Davidson, Gull, Ishibashi  \&
  Hillier}{Smith et~al.}{2003b}]{Smith2003LatitudedependentCarinae}
Smith N.,  Davidson K.,  Gull T.~R.,  Ishibashi K.,   Hillier D.~J.,  2003b,
  \mn@doi [The Astrophysical Journal] {10.1086/367641}, 586, 432

\bibitem[\protect\citeauthoryear{Steiner \& Damineli}{Steiner \&
  Damineli}{2004}]{Steiner2004DetectionCarinae}
Steiner J.~E.,  Damineli A.,  2004, \mn@doi [The Astrophysical Journal]
  {10.1086/424831}, 612, 133

\bibitem[\protect\citeauthoryear{Stevens \& Pollock}{Stevens \&
  Pollock}{1994}]{Stevens1994Stagnation-pointSystems}
Stevens I.~R.,  Pollock A. M.~T.,  1994, \mn@doi [Monthly Notices of the Royal
  Astronomical Society] {10.1093/mnras/269.2.226}, 269, 226

\bibitem[\protect\citeauthoryear{Stickland, Bromage, Budding, Burton  \&
  Willis}{Stickland et~al.}{1984}]{Stickland1984UltravioletCephei}
Stickland D.~J.,  Bromage G.~E.,  Budding E.,  Burton W.~M.,   Willis A.~J.,
  1984, Astronomy {\&} Astrophysics, 134, 37

\bibitem[\protect\citeauthoryear{Teodoro et~al.,}{Teodoro
  et~al.}{2012}]{Teodoro2012HePassage}
Teodoro M.,  et~al., 2012, \mn@doi [Astrophysical Journal]
  {10.1088/0004-637X/746/1/73}, 746, 73

\bibitem[\protect\citeauthoryear{Teodoro et~al.,}{Teodoro
  et~al.}{2016}]{Teodoro2016HeMINIMA}
Teodoro M.,  et~al., 2016, \mn@doi [The Astrophysical Journal]
  {10.3847/0004-637x/819/2/131}, 819, 47

\bibitem[\protect\citeauthoryear{Van Der~Walt, Colbert  \& Varoquaux}{Van
  Der~Walt et~al.}{2011}]{VanDerWalt2011TheComputation}
Van Der~Walt S.,  Colbert S.~C.,   Varoquaux G.,  2011, \mn@doi [Computing in
  Science and Engineering] {10.1109/MCSE.2011.37}, 13, 22

\bibitem[\protect\citeauthoryear{Walborn}{Walborn}{1973}]{Walborn1973SomeComplex}
Walborn N.~R.,  1973, \mn@doi [The Astrophysical Journal] {10.1086/151891},
  179, 517

\bibitem[\protect\citeauthoryear{Walborn \& Liller}{Walborn \&
  Liller}{1977}]{Walborn1977TheNebula}
Walborn N.~R.,  Liller M.~H.,  1977, \mn@doi [The Astrophysical Journal]
  {10.1086/154917}, 211, 181

\bibitem[\protect\citeauthoryear{Walder}{Walder}{1995}]{Walder1995SIMULATIONS}
Walder R.,  1995, in Hucht K. A. v.~d.,  Williams P.,  eds, WR stars: binaries,
  colliding winds, evolution.. IAU Symposium No. 163, pp 420--424

\bibitem[\protect\citeauthoryear{Weigelt \& Ebersberger}{Weigelt \&
  Ebersberger}{1986}]{Weigelt1986EtaInterferometry}
Weigelt G.,  Ebersberger J.,  1986, Astronomy {\&} Astrophysics, 163, L5

\bibitem[\protect\citeauthoryear{Willis}{Willis}{1982}]{Willis1982P-CygniStars}
Willis A.~J.,  1982, \mn@doi [Monthly Notices of the Royal Astronomical
  Society] {10.1093/mnras/198.4.897}, 198, 897

\bibitem[\protect\citeauthoryear{Wu, Smith, Close, Males  \& Morzinski}{Wu
  et~al.}{2017}]{Wu2017ResolvingCarinae}
Wu Y.-L.,  Smith N.,  Close L.~M.,  Males J.~R.,   Morzinski K.~M.,  2017,
  \mn@doi [The Astrophysical Journal Letters] {10.3847/2041-8213/aa70ed}, 841,
  L7

\bibitem[\protect\citeauthoryear{de Vaucouleurs \& Eggen}{de~Vaucouleurs \&
  Eggen}{1952}]{deVaucouleurs1952TheCarinae}
de Vaucouleurs G.,  Eggen O.~J.,  1952, \mn@doi [Publications of the
  Astronomical Society of the Pacific] {10.1086/126457}, 64, 185

\makeatother
\end{thebibliography}



\appendix

\section{Keplerian radial velocity curves}
\label{sec:keplerian_radial_velocity_curves}

\begin{figure}
	\includegraphics[width=1\columnwidth]{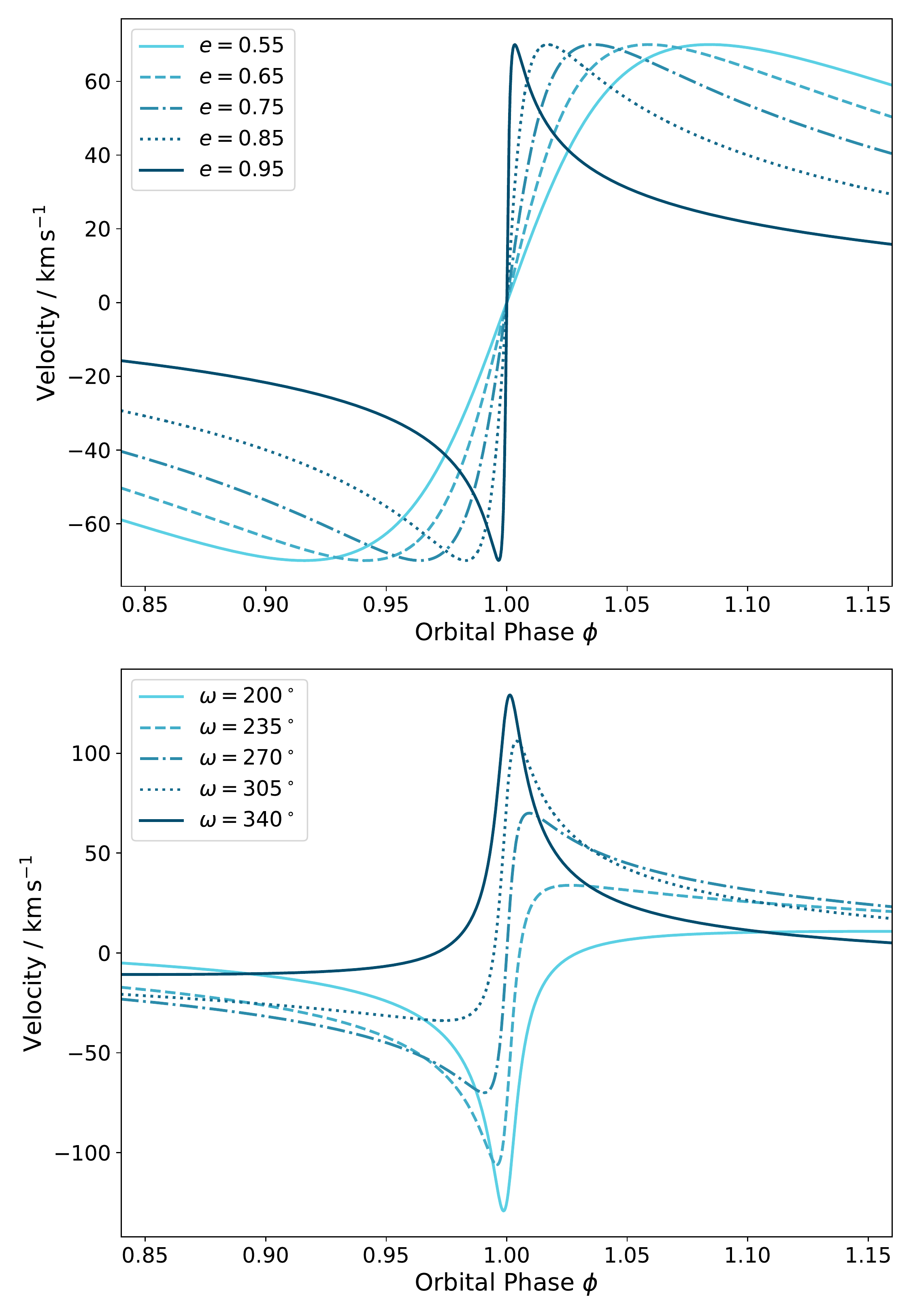}
    \caption{Examples of Keplerian radial velocity curves for changes to the orbital parameters eccentricity, $e$, and longitude of periastron, $\omega$.}
    \label{fig:keplerian_solutions_testing}
\end{figure}

We present a series of Keplerian radial velocity curves showing how the line-of-sight velocities depend on the eccentricity, $e$, and longitude of periastron, $\omega$ (see Figure \ref{fig:keplerian_solutions_testing}).

First, we test the dependence on $e$ over the interval $0.55 < e < 0.95$, for a fixed $\omega=270^\circ$. As the eccentricity increases the radial velocities show faster changes around periastron. Second, we test the dependence on $\omega$ over the interval $200^\circ < \omega < 340^\circ$, for a fixed $e=0.9$. For $\omega=270^\circ$ we view the system directly from the apastron side. As $\omega$ increases (decreases) from this value we see a corresponding increases in positive (negative) velocities around periastron. This is a result of the faster orbital velocities at periastron having a greater magnitude when projected onto these line-of-sights.

\section{The effect of the inner boundary}
\label{sec:inner_boundary_testing}

\begin{figure}
	\includegraphics[width=1\columnwidth]{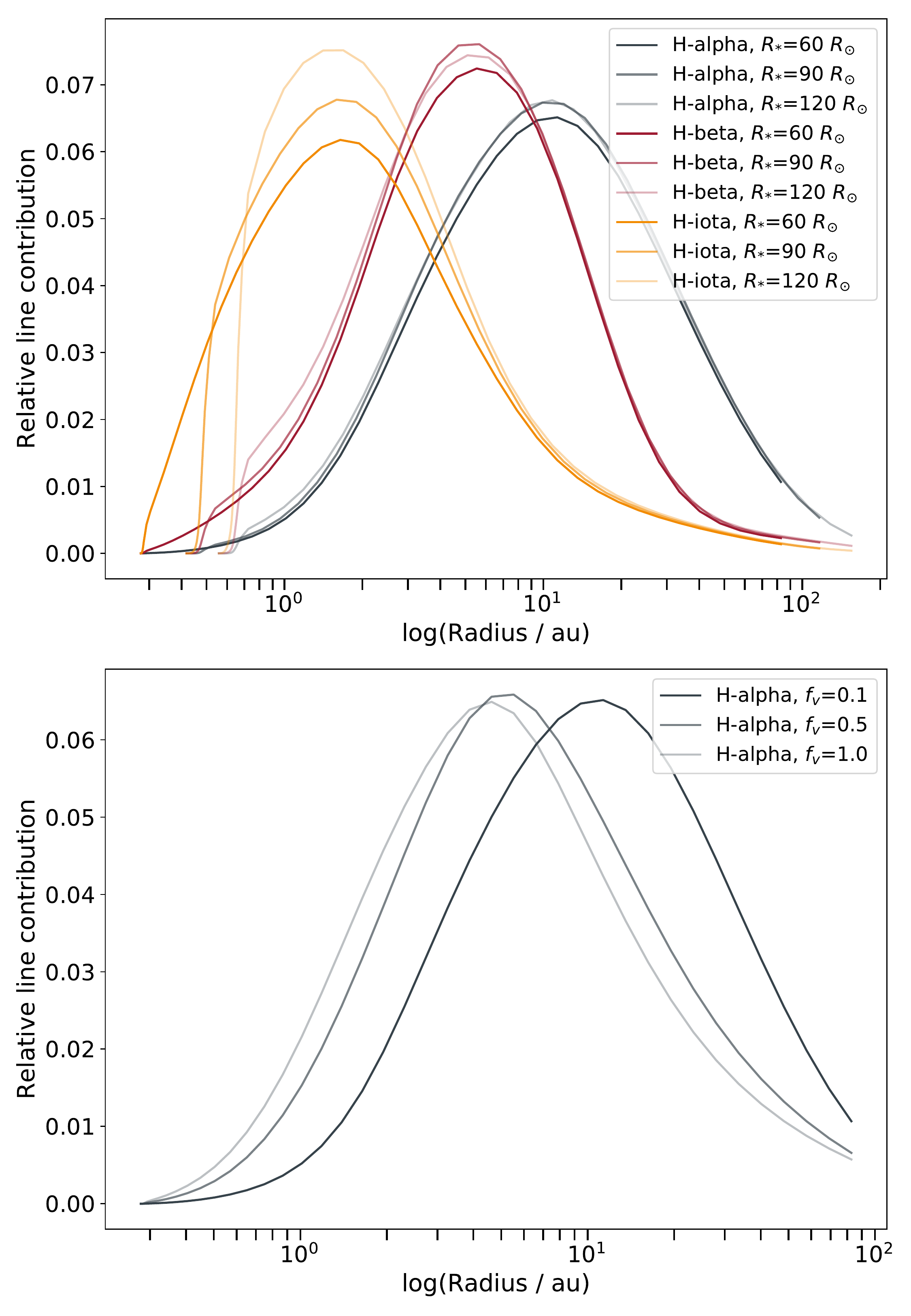}
    \caption{Top panel: a comparison of the relative line contributions for lines H-alpha (black), H-beta (red) and H-iota (orange) for different \texttt{CMFGEN} inner boundaries (darker lines indicate smaller radii). Bottom panel: a comparison of the relative line contributions for H-alpha for different \texttt{CMFGEN} volume filling factors. (darker lines indicate more clumping)}
    \label{fig:testing_cmfgen}
\end{figure}

\begin{figure}
	\includegraphics[width=1\columnwidth]{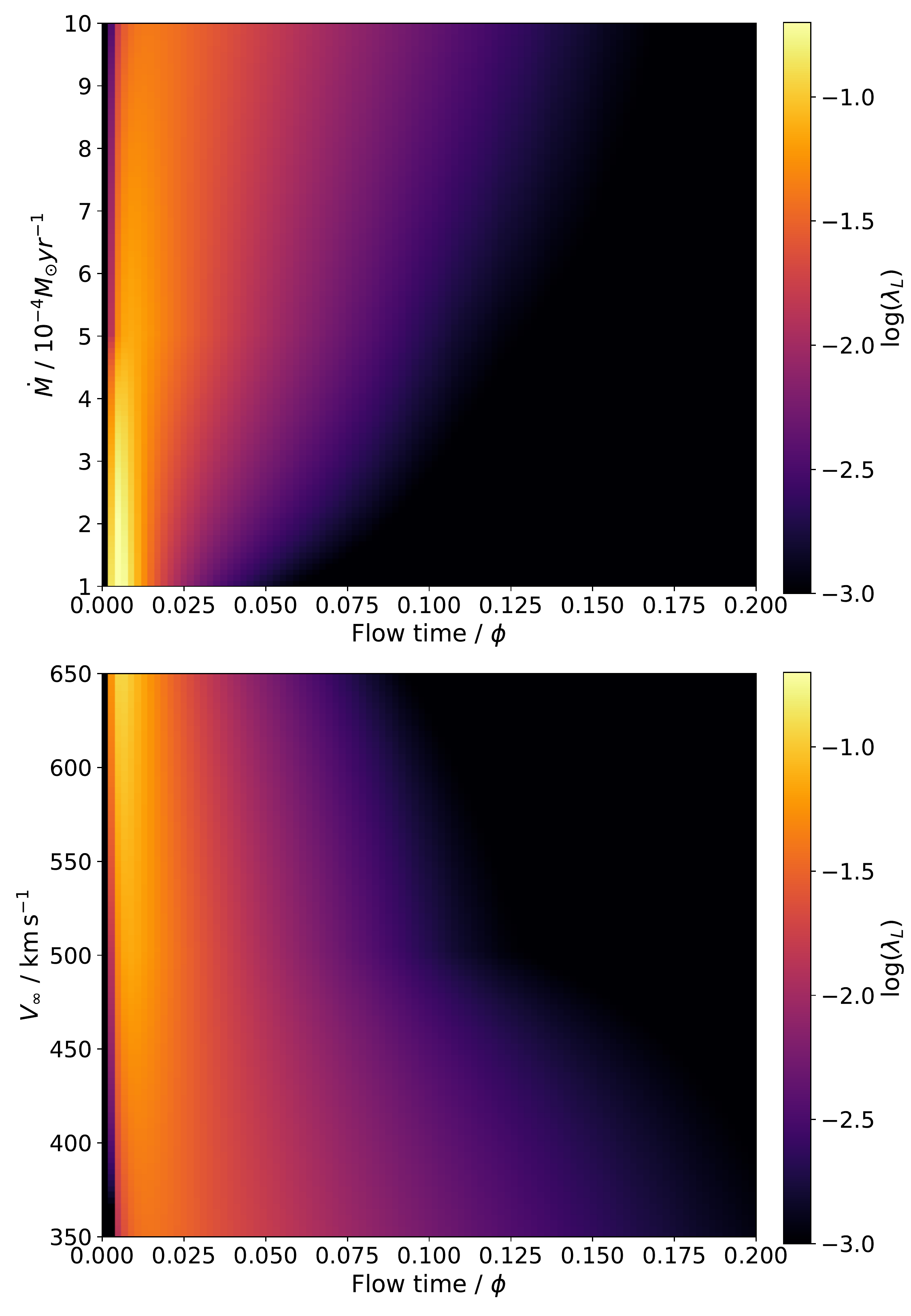}
    \caption{Line-formation kernels for an interpolated 3x3 grid of \texttt{CMFGEN} simulations varying $\Dot{M}$ and $v_{\infty}$. Each kernel is comprised of the linear luminosity density, $\lambda_{L,n}$, as a function of wind flow time.}
    \label{fig:rt_interp_grid_slices}
\end{figure}

We test the effect of changing the inner boundary, $R_{*}$, on the resulting line-formation regions in our \texttt{CMFGEN} radiative transfer simulations. We run the simulation in a high mass-loss regime, similar to the conditions found in stars such as \etacar{}. The input parameters are luminosity, $L=5 \times 10^6 \solarl{}$, mass-loss rate, $\Dot{M}=5 \times 10^{-4} \masslossrate{}$, terminal wind velocity, $v_{\infty}=500 \kmpers{}$, and volume filling factor, $f_{\rm{v}}=0.1$. We test $R_{*}$ for values of $60 \rm{R_{\odot}}$, $90 \rm{R_{\odot}}$ and $120 \rm{R_{\odot}}$.

The results are presented in the top panel of Figure \ref{fig:testing_cmfgen}. The relative line contribution, $\xi$, is related to the equivalent width (EW) of a line by:
\begin{equation}
EW = \int_{R_{*}}^{\infty} \xi(r) d(\log(r)).
	\label{eq:line_contribution_equivalent_width}
\end{equation}
As previously found by \citet[][]{Hillier2001} the extreme mass loss prevents the determination of $R_{*}$ accurately within our simulations as the results appear insensitive to this parameter. We set $R_{*} = 60 \rm{R_{\odot}}$ as the value for all further model runs. 

\section{The effect of clumping}
\label{sec:clumping_testing}

We test the effect of changing the clumping on the resulting line-formation regions in our \texttt{CMFGEN} radiative transfer simulations. The standard way of including this effect in stellar atmosphere codes is to introduce optically thin clumps within a void inter-clump medium. The over-density inside the clumps is determined by a so-called clumping factor, $f_{\rm{cl}}$, or volume filling factor $f_{\rm{v}}$ defined by
\begin{align}
    \label{eq:clumping factor}
& f_{\rm{cl}} = \langle \rho^{2} \rangle / \langle \rho \rangle ^{2}, \\
	\label{eq:volume filling factor}
& f_{\rm{v}} = {f_{\rm{cl}}}^{-1},
\end{align}
where $\rho$ is the density and the angle brackets denote volume averaging. We run the simulation with input parameters of luminosity, $L=5 \times 10^6 \solarl{}$, mass-loss rate, $\Dot{M}=5 \times 10^{-4} \masslossrate{}$, terminal wind velocity, $v_{\infty}=500 \kmpers{}$, and stellar radius, $R_{*} = 60 \rm{R_{\odot}}$. We test $f_{\rm{v}}$ for values of $0.1$, $0.5$ and $1.0$.

The results are presented in the bottom panel of Figure \ref{fig:testing_cmfgen} and the relative line contribution is defined in Equation \ref{eq:line_contribution_equivalent_width}. We find that a clumpy wind creates higher densities at larger radii; and therefore, line emission is possible at more extended regions from the star than previously allowed in a homogeneous wind. We set $f_{\rm{v}} = 0.1$ as the value for all further model runs.

\section{Line-formation kernel degeneracy}
\label{sec:line_formation_kernel_degeneracy}

We test the effect of changing the mass-loss rate, $\Dot{M}$, and wind terminal velocity, $v_{\infty}$, on the resulting line-formation kernels in our \texttt{CMFGEN} radiative transfer simulations. We setup the simulation with input parameters of luminosity, $L=5 \times 10^6 \solarl{}$, mass-loss rate, $\Dot{M}=5 \times 10^{-4} \masslossrate{}$, terminal wind velocity, $v_{\infty}=500 \kmpers{}$, stellar radius, $R_{*} = 60 \rm{R_{\odot}}$ and volume filling factor, $f_{\rm{v}}=0.1$. We run a coarse 3x3 grid of simulations varying $\Dot{M}$ and $v_{\infty}$.

The results are presented in Figure \ref{fig:rt_interp_grid_slices}. We show the line-formation kernels -- linear luminosity density, $\lambda_{L,n}$, as a function of wind flow time -- interpolated between grid runs. In the top panel $v_{\infty}$ is held constant at $500 \kmpers{}$ and in the bottom panel $\Dot{M}$ is held constant at $5 \times 10^{-4} \masslossrate{}$. We find line formation occurs at larger radii as $\Dot{M}$ increases and occurs at lower radii as $v_{\infty}$ increases. These two parameters are highly degenerate.


\bsp	
\label{lastpage}
\end{document}